\documentclass[review]{elsarticle}
\makeatletter                        
\def\ps@pprintTitle{%
 \let\@oddhead\@empty
 \let\@evenhead\@empty
 \def\@oddfoot{}%
 \let\@evenfoot\@oddfoot}
\makeatother
\usepackage{lineno,hyperref,graphicx}
\usepackage[normalem]{ulem}
\usepackage[margin=1.0in]{geometry} 
\modulolinenumbers[1]
\setpagewiselinenumbers
\usepackage{multirow}
\usepackage{subfigure}
\usepackage{amsfonts}
\usepackage{amsmath}
\usepackage{booktabs, threeparttable}
\usepackage{array}
\newcolumntype{L}[1]{>{\raggedright\arraybackslash}p{#1}}

\usepackage{color}

\begin{document}

\title{Dissipation and adhesion hysteresis between $(010)$ forsterite surfaces using molecular-dynamics simulation and the Jarzynski equality}

\author[1]{Baochi Doan}
\author[2,3,4]{Patrick K. Schelling\corref{mycorrespondingauthor}}
\cortext[mycorrespondingauthor]{Corresponding author.}
\ead{patrick.schelling@ucf.edu}
\address[1] {Department of Materials Science and Engineering, University of Central Florida, FL 32816-2385, USA}
\address[2]{Department of Physics, University of Central Florida, Orlando, FL 32816-2385, USA }
 \address[3] { Advanced Materials Processing and Analysis Center, University of Central Florida, Orlando, FL 32816-2385, USA }
 \address[4]   {  Renewable Energy and Chemical Transformations (REACT) Cluster, University of Central Florida,  University of Central Florida, Orlando, FL 32816-2385, USA}

\begin{abstract}
Dissipation and adhesion are important in many areas of materials science, including friction and lubrication, cold spray deposition, and micro-electromechanical systems (MEMS). Another interesting problem is the adhesion of mineral grains during the early stages of planetesimal formation in the early solar system.
Molecular-dynamics (MD) simulation has often been used to elucidate dissipative properties, most often in the simulation of sliding friction. In this paper, we
demonstrate how the reversible and irreversible work associated with interactions between planar surfaces can be calculated using the dynamical contact simulation approach based on MD and empirical potentials. Moreover, it is demonstrated how the approach can obtain the free-energy $\Delta A(z)$ as a function of separation between two slabs using the Jarzynksi equality applied to an ensemble of trajectories which deviate significantly from equilibrium. Furthermore, the dissipative work can also be obtained using this method without the need to compute an entire cycle from approach to retraction. It is expected that this technique might be used to efficiently compute dissipative properties which might enable the use of more accurate approaches including density-functional theory. In this paper, we present results obtained for forsterite surfaces both with and without MgO-vacancy surface defects. It is shown that strong dissipation is possible when MgO-vacancy defects are present. The mechanism for strong dissipation is connected to the tendency of less strongly-bound surface units to undergo large displacements including mass transfer between the two surfaces. Systems with strong dissipation tend to exhibit a long-tailed distribution rather than the Gaussian distribution often anticipated in near-equilibrium applications of the JE. 

\end{abstract}

\maketitle

\section{Introduction}

Dissipation and adhesion are important in a wide array of phenomena. For example, in additive manufacturing, cold-spray technology uses a supersonic gas jet to accelerate micron-sized powders which form a coating on impacting a substrate\cite{CHAMPAGNE2018799}. In this process, high particle impact velocities result in plastic deformation and strong adhesion. In microelectromechanical systems (MEMS), stiction is another important consideration for both fabrication and device operation\cite{KRIM2002741,Zhuang_2005}. Another interesting problem is the early stages of planet formation, where  the adhesion of micrometer-sized dust grains is an essential step which is still rather poorly understood. Specifically, experimental evidence suggests that aggregates can grow to millimeter scales, but beyond these length scales collisions tend to result in destruction of aggregates rather than further growth\cite{blum:2000,blum:2008,blum:2010,blum:2018}. In each of these problems or applications, the central goal is to understand dissipation and adhesion mechanisms beginning with atomic-scale interactions at interfaces.

Theoretical methods initially comprised simple, phenomenological models. One of the most important early models of adhesion includes the Johnson-Kendall-Roberts model\cite{jkr:1971}. One basic feature of this model is adhesion hysteresis, whereby on approach surfaces are not in direct contact, whereas in the rebound phase a ``neck'' is formed which requires a finite tensile force to break. Sometimes it is said that there is a ``jump-in'' to contact, and a ``jump-out'' at a finite tensile pull-off force \cite{Ciavarella_2019}. As such this model describes a process which is inherently irreversible and dissipative. Adhesion hysteresis has been studied in a number of different contexts including modeling of atomic-force microscopy (AFM) experiments. In the case of colliding particles, dissipation of the incident kinetic energy is required to prevent breaking the adhesive neck. In the description of sliding friction, the Prandtl-Tomlinson model\cite{Tomlinson:1929,Tomanek:1991,Popov:2010} is a standard model used to describe the physics of stick-slip friction, whereas a sliding at an interface results in sticking until the lateral force is large enough to activate bond breakage. During the activated process, the frictional force is greatly reduced, which can be observed as a sawtooth modulation in a friction-force microscope\cite{Bennewitz_2014}.

MD simulation of atomic-force microscopy (AFM) operation has proven useful for prediction of dissipation and local elastic properties\cite{TREVETHAN2003497,Trevethan:2004,Kim_2012,Onofrio_2013}. Here we describe the dynamical contact simulation (DCS) approach developed previously\cite{Kim_2012}. The basic idea is to impart an initial downward (i.e. towards the substrate) velocity $-v_{0}$ ($v_{0}>0$) while imposing a constant restoring force $F_{res}$ ($F_{res}>0$). With a suitable choice of $-v_{0}$ and $F_{res}$, the tip approaches the surface but does not stick. By integrating the work done by the interaction force, the dissipative work can be determined. Specifically, by starting at a height $z_{0}$ above the substrate, the dissipated work is given by,
\begin{equation} \label{wdissint}
W_{diss}=-W_{int}=-\int_{z_{0}}^{z_{c}} F_{int}^{(app)}(z) dz + \int_{z_{c}}^{z_{0}} F_{int}^{(ret)}(z)dz
\end{equation}
in which $z_{c}$ is the distance of closest approach which depends on the choice of $v_{0}$ and $F_{res}$, $F_{int}^{(app)}(z)$ is the interaction force during approach, and $F_{int}^{(ret)}(z)$  is the force during the retraction phase. Note that the restoring force $F_{res}$ does no work during a closed cycle. The dissipative work is due entirely to the adhesion hysteresis. Specifically, for a given value of $z$ the interaction force on approach and retraction are generally expected to be different.

We will consider the interaction between two atomically smooth slabs of material separated by a distance $z$. The DCS method is applied to compute the adhesive and dissipative properties of the interface. 
If the approach and retraction phases were completely reversible and isothermal, then the interaction force $F_{int}(z)$ between the slabs would be related to the Helmholtz free energy $A(z)$, specifically,
\begin{equation}
F_{int} = -\frac{dA}{dz}
\end{equation}
in which $z$ represents the separation between the surfaces. However, because approach and retraction happen at finite velocity, the two slabs are always out of equilibrium, and any approach and retraction will be irreversible and hence dissipative. For strong enough dissipation, matter transfer between the slabs or even sticking behavior might occur. A significant obstacle to understanding the dissipation and adhesion is that the function $A(z)$ is not known. Therefore, while the interaction force between two surfaces can be computed, direct integration cannot be used directly to determine $A(z)$. In short, some portion of the integrated work is dissipative and irreversible, but apart from analyzing a closed cycle of approach and retraction it is currently difficult to directly determine dissipative work.

In this paper, we will apply the Jarzynski equality (JE)\cite{Jarzynski:1997uj}  to compute both free-energy differences $\Delta A(z)$ and also the dissipative work. Typically, the JE has been used to compute free-energy differences $\Delta A$ from an ensemble of calculations potentially arbitrarily far from equilibrium. Generally, dissipative properties have been examined to elucidate statistical and numerical aspects related to convergence of the JE. More recently, however, the JE has been used to compute not only free-energy differences but also dissipative work. For example, internal friction was studied using a simple spring-dashpot model of a polymer\cite{Kailasham:2020wd} and the JE. In the next section we detail some aspects of the JE for calculation of free-energy curves $\Delta A(z)$ for interaction surfaces including dissipative work.

\section{The Jarzynski Equality}

Several breakthroughs have been made in understanding how reversible thermodynamics, including free energy differences $\Delta A$, can be obtained from either experiments or computational results carried out in non-equilibrium conditions. It was shown by Jarzynski\cite{Jarzynski:1997uj} that equilibrium free-energy differences can be determined from an ensemble of trajectories which are not in equilibrium. Specifically, the Jarzynski equality (JE) states
\begin{equation}
\exp \left(-\beta \Delta A\right) = \langle \exp \left(-\beta W\right)\rangle
\end{equation}
in which $\beta = \frac{1 }{k_{B}T}$, $\Delta A$ is the free-energy difference between two equilibrium states, and $W$ is the work done by an external force. The angle brackets indicate an average over an ensemble of trajectories. Alternately one can write for the free-energy difference,
\begin{equation} \label{free}
\Delta A = -\beta^{-1} \ln \langle \exp\left(-\beta W\right) \rangle
\end{equation}
It is also clear that $\langle W \rangle > \Delta A$ and that the dissipative or irreversible work is given by $W_{diss}=\langle W \rangle - \Delta A$. Often these quantities are used to assess the ability of the JE to determine $\Delta A$, but more recently using the JE to determine dissipative work has been reported\cite{Kailasham:2020wd}.

More detailed information about the free-energy curve and dissipation on approach and retraction, rather than just for a closed cycle, can be obtained using the JE. The simulation approach used here generally follows the DCS method\cite{Kim_2012}.
We consider two atomically-flat slabs initially separated by a distance $z_{0}$ large enough so that their mutual interactions are small. These slabs are in thermal equilibrium at the same initial temperature maintained by
contact with separate reservoirs. The two slabs are them given an initial kinetic energy $W_{i}=K_{i}=mv_{0}^{2}$, in which $m$ is the (identical) mass of each slab and $v_{0}$ is the speed of each slab in the 
center-of-mass reference frame. The direction of motion is chosen so that the slab separation $z$ decreases. The slabs are also acted on by a constant, uniform restoring force $F_{res}$ which prevents the slabs from sticking together. The restoring forces are applied evenly to each ion comprising the two slabs. The work done by the restoring force in moving from separation $z_{0}$ to separation $z$ is given by $W_{res}(z)=F_{res}(z-z_{0})$. 
When the slabs are at separation $z$, their kinetic energy associated with translational motion is $K(z)$. This motion could be stopped by performing work $W(z)=-K(z)$. These contributions
to the net work $W_{net}$ are summed,
\begin{equation}
 W_{net}(z)=K_{i}+W_{res}(z)+W(z) = W_{0}(z)-K(z)
\end{equation}
in which $W_{0}(z)=K_{i}+W_{res}(z)$ is the net reversible work due to the restoring force and to given the slabs the initial kinetic energy $K_{i}$.
Then the JE can be applied to obtain the free-energy difference,
\begin{equation} \label{JE}
A(z)-A(z_{0}) = -\frac{1}{\beta}  \ln \langle \exp \{ -\beta \left[W_{0}(z)-K(z)\right ] \} \rangle 
\end{equation}
Because $K(z)$ is usually significantly greater than $k_{B}T$, it is essential to retain $W_{0}(z)$ within the exponential during averaging. 
The averaging in Eq. \ref{JE} is called the Jarzynski estimator\cite{Gore:2003vb}. Other approaches include the fluctuation-dissipation estimator and cumulant expansions to higher-order terms\cite{Gore:2003vb}. It has previously been shown that the Jarzynski estimator is superior to these other approaches\cite{Gore:2003vb}. In near equilibrium cases, systematic bias in the Jarzynski estimator has been previously quantified\cite{Gore:2003vb}. In the results presented here, results are obtained in a regime of strong dissipation where the system tends to be very far from equilibrium during the simulation.

Finally, the average dissipative work $ W_{diss}(z)$ as a function of $z$ is obtained from,
\begin{equation} \label{wdissnotclosed}
W_{diss}(z)=W_{0}(z) -\langle K \rangle_{z} - \left[A(z)-A(z_{0}) \right] 
\end{equation}
In the case of a reversible process, the dissipative work is zero. Then one has the equality $\langle K \rangle_{z} = W_{0}(z)-\left[A(z)-A(z_{0}) \right] $.

In the dynamical contact simulation (DCS) approach\cite{Kim_2012}, dissipative work can be obtained only from direct integration of the interaction work for a closed path. Equivalently,  direct calculation of $W_{0}(z_{0}) -\langle K(z_{0}) \rangle$ for a closed path with final separation $z=z_{0}$ corresponds to the dissipative work. Specifically, assuming there is no damage to the slabs a closed path corresponds to 
$\Delta A=0$ and then,
\begin{equation} \label{wdiss2}
W_{diss}(z_{0})=W_{0}(z_{0}) -\langle K \rangle_{z_{0}}
\end{equation}
in which $z_{0}$ is the slab separation after the retraction phase. Since $z_{0}$ is also the slab separation in the initial state, Eq. \ref{wdiss2} represents a calculation over a closed path including the approach and retraction phases. In addition, the expression in Eq. \ref{wdiss2} is equivalent to the integrated work in Eq. \ref{wdissint}. Hence we can make the association $-W_{int} = W_{diss}(z_{0})$ when $W_{diss}(z_{0})$ is obtained during the retraction phase.

The advantage of using the JE is apparent in the above expressions. Specifically, while $W_{diss}(z)$ (or $-W_{int}$) can be obtained for a closed path (i.e. $z=z_{0}$) either from Eq. \ref{wdissint} or equivalently Eq. \ref{wdiss2}, it is not generally possible to find $W_{diss}(z)$ for any position $z$. This is because generally the free energy function $A(z)$ is not known. By determining $\Delta A(z)$ using the JE, $W_{diss}(z)$ can be revealed, which enables calculation and elucidation of dissipation mechanisms at any point along a trajectory, and also possibly could eliminate the costly requirement of computing closed cycles. Specifically, if $W_{diss}(z)$ can be determined for smaller segments of a trajectory, it would be more feasible to introduce accurate approaches including density-functional theory (DFT) to capture chemical reactions as a mechanism for dissipation.

\section{Computational approach}

\begin{figure}  \label{figure1}
\begin{center}
 \includegraphics[scale=0.20]{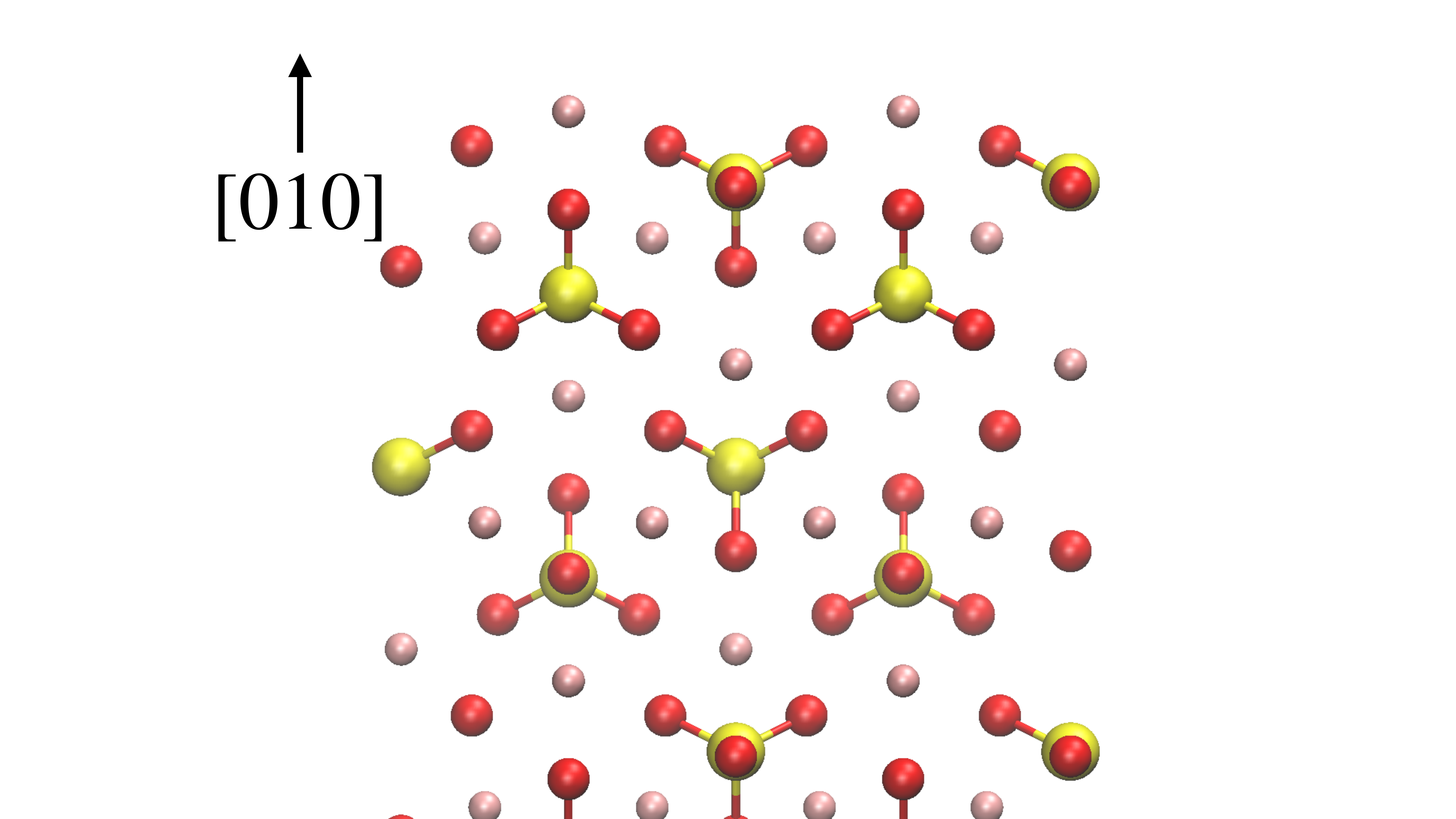}
\end{center}
\caption{
The simulated $(010)$ surface of forsterite. Pink spheres are Mg, yellow spheres are Si, and red spheres represent O ions. The $[100]$ direction is out of the page.}
\end{figure}

The LAMMPS simulation code was used for all calculations\cite{plimpton:1995}. The real-space Wolf method for Coulomb summations was used\cite{Wolf_1999} with parameters $\alpha= 0.2 \AA^{-1} $ and $R_{c}=12\AA$. These parameters were tested for a range of minerals including forsterite, fayalite, and pyroxenes in simulations previously reported\cite{quadery:2015}. In addition to Coulombic interactions, the model consists of pairwise interaction terms with the parameterization reported previously\cite{quadery:2015}. The pairwise interaction is given by the potential energy function $u_{ij}$ for pair $ij$,
\begin{equation} \label{potential}
u_{ij}(r_{ij}) =   \frac{q_{i}q_{j}}{r_{ij}} +D_{ij} \left[ \left( 1 - e^{-a_{ij}(r_{ij} -r_{0})}\right)^{2}-1 \right] + \frac{C_{ij}}{{r_{ij}}^{12}}
\end{equation}
in which $q_{i}$ and $q_{j}$ represent ionic charges, and the pair separation distance is given by $r_{ij}$. The parameters used for the pair potentials are given in Table \ref{table1}. Bulk forsterite Mg$_2$SiO$_4$ has an orthorhombic Bravais lattice and is characterized by space group Pbnm.
At $T=0K$, the computed relaxed lattice parameters were $a=4.781\AA$, $b=10.142\AA$, and $c=5.963\AA$, in very good agreement with experiment\cite{quadery:2015}. The MD time step was taken to be $0.5$fs.  Simulation of bulk forsterite at $T=500$K and constant pressure $p=0$ resulted in lattice parameters $a=4.812\AA$, $b=10.210\AA$, and $c=6.0034\AA$. These values were used in subsequent constant-volume simulations at $T=500$K.

The structures simulated consisted of $(010)$ forsterite surfaces. The structure of the surface used is shown in Fig. \ref{figure1}. The surface termination was chosen such that the simulated slabs were non-polar. The structure was identical to previously reported simulations of water dissociation at forsterite\cite{de_Leeuw_2000}. The surface energy $\gamma_{(010)}$ of the slab was determined from the expression,
\begin{equation}
\gamma_{(010)} = \frac{E_{slab} - n E_{bulk}}{2S}
\label{surfaceenergy}
\end{equation}
in which $S$ is surface area, $E_{slab}$ is the energy of the slab with $n$ unit cells, and $E_{bulk}$ is the energy per unit cell in a bulk periodic structure. For slab calculations, periodic-boundary conditions were applied only in the directions perpendicular to the slab normal direction. The dimensions of the simulation cell for the periodic $[100]$ and $[001]$ directions were fixed by the lattice parameters $a$ and $c$ of the bulk lattice at $T=500$K. The slab simulated consisted of $6$ unit cells along $[100]$ and $5$ units along $[001]$. The slab was $3$ unit cells thick along $[010]$. For these dimensions, one simulated $(010)$ slab was comprised of $90$ unit cells and $360$ Mg$_2$SiO$_4$ formula units. The resulting value of $\gamma_{(010)} = 0.084 eV \AA^{-2}$ at $T=0$K and using the $T=0$K lattice parameters is in essentially exact agreement with the value reported previously\cite{de_Leeuw_2000}, despite the use of a different empirical potential.

\begin{table}
\begin{center}
\caption{Potential parameters for forsterite model first reported in \citep{quadery:2015}. The charges in the model are $q_{Mg}=1.20|e|$, $q_{Si}=2.40|e|$, and $q_{O}=-1.20|e|$. For Mg-Mg pairs, only a repulsive Coulomb interaction is included. Parameters correspond to those shown in Eq. \ref{potential} }
\begin{tabular} {|c|c|c|c|c|}
\hline
Pair & $D$(eV) & $a(\AA^{-1})$ & $r_{0}$ ($\AA$) & $C$ (eV$\AA^{12}$) \\\hline 
Mg-O & 0.123583   &  2.045583 &  2.424824  &  5.0 \\\hline
Si-O   & 0.443427   &  1.758024 &  2.081625  &  1.0 \\\hline
O-O   & 0.042323   & 1.311417  &  3.762599   & 22.0\\\hline
\end{tabular}
\label{table1}
\end{center}
\end{table}

For DCS calculations, two different structures were used corresponding to different surface area $S$. In the larger structure, both $(010)$ slab structures had $6 \times 5$ unit cells along the $[100]$ and $[001]$ directions respectively. In the smaller structure, both surface had $2 \times 2$ unit cells along $[100]$ and $[001]$. In both cases the slabs were $3$ unit cells in thickness. Because $b=10.210\AA$, the thickness of the slabs is much greater than the cutoff distance for the pair potential, and hence we do not expect the results to depend significantly on slab thickness. For the larger slabs, the surface area is about $S \approx 866.8 \AA^2$, and for the smaller slabs $S \approx 115.6 \AA^2$. 

The main reason for including structures with smaller surface areas was to address the well-known difficulty with the JE which occurs when statistical fluctuations in the 
work ensemble become significantly greater than $k_{B}T$\cite{Jarzynski:1997uj}. Averaging of the Jarzynski estimator is substantially more accurate when some ensemble members have zero or even negative dissipation\cite{Gore:2003vb}. This situation is more difficult to realize for larger systems. In simulating a smaller surface area, only fluctuations related to the simulation supercell are relevant since periodic images are identical. Therefore, work distributions in the following are defined with respect to a single periodic unit. When simulating a larger periodic unit, fluctuations are larger in magnitude, and it is less likely that ensemble members will have zero or even negative dissipation, and hence evaluation of the Jarzynski estimator is more prone to statistical errors. Another drawback of the large surface area structures was computational time which present a practical limit for the number of ensemble members. For simulations with $S \approx 115.6 \AA^2$, $N=500$ members were included in each ensemble. By contrast, calculations with area $S \approx 866.8 \AA^2$ included only $N=200$ members were simulated in each ensemble. 

To generate two opposing surfaces, two slabs are translated a distance $z=6\AA$ apart from the location where they would form a perfect crystal interface. In addition to translation, one of the slabs is rotated by $180^{\circ}$ about a $[010]$ axis. In this way, the two opposing surfaces do not form a perfect crystal if brought together. Rather, the interface generated is comparable to a twist-grain boundary. However, in the following DCS calculations, the surfaces generally do not come close enough to adhere and form a grain-boundary structure. While interactions are not zero at $z=6\AA$, they are very small as the following calculations demonstrate. This is not surprising since the slabs were non polar, and hence Coulomb interactions are expected to decrease rapidly with increasing $z$. The choice of $z=6\AA$ rather than a separation greater that the cutoff distance was made to decrease computational cost. Moreover, since $\Delta A$ is an equilibrium free energy difference, calculation of this quantity should not depend on the initial starting point.

Defective structures were generated by removal of neutral MgO units from the opposing surfaces. It has been found previously that MgO Schottky defects represent a relatively low-energy defect structure in olivine\cite{Demouchy:2021,brodholt:1997}. For the small slabs with area $S=115.6\AA^{2}$, both opposing surfaces were deficient in 6 Mg and 6 O ions taken from the outermost layer. For larger slabs with area $S \approx 866.8 \AA^2$, 10 Mg and 10 O ions were removed from the outermost layer. Due to the large difference in surface area, the defect density was significantly greater for the small slabs. Specifically, the surface density $n_{vac}$ of ion vacancies was $n_{vac}=0.052 \AA^{-2}$ for small slabs, and $n_{vac}=0.011 \AA^{-2}$ for larger slabs, thereby differing by a factor of nearly $5$. Defective slabs were neutral and also did not possess a net dipole moment. In contrast to structures without defects, MgO-vacancy defects lead to slabs which are not atomically flat.

In the following, physical quantities, including $\Delta A$, $W_{int}$, and even $F_{res}$ are scaled by the surface area to enable comparison between the two different system sizes. Specifically, we define the restoring stress $\sigma_{res}=F_{res}/S$ and interaction stress $\sigma_{int} = \frac{F_{int}}{S}$. Similarly, the interaction work per unit area is defined as $\gamma_{int} = W_{int}/S$, and the dissipative work per unit area $\gamma_{diss}=W_{diss}/S$. The initial kinetic energy associated with a slab can also be normalized by the surface area $K_{i}/S$. Specific quantities of this kind are more fundamental for understanding the process, and moreover enable comparison between simulations with different surface areas $S$. Unless otherwise stated, reported quantities are averaged over the ensemble generated with independent initial conditions.

As a first step, structures were equilibrated to $T=500K$ using the Nose-Hoover thermostat\cite{Nos__1984,Hoover:1985wf}. Two slabs were placed at a separation of $6 \AA$ and then were equilibrated over a time of $50$ps. In the case of slabs with MgO-vacancy defects, the separation between the centers of mass of the two slabs was slightly larger due to the removal of ions from the opposing surfaces. Specifically, $z_{0}=6.47\AA$ was obtained for the initial separation of the defective small slabs, and $z_0=6.1\AA$ for large defective slabs. A separate thermostat was applied to both slabs. During equilibration, a constraint force was uniformly applied to the ions to maintain the initial separation $z_{0}$. After equilibration, the atoms in each slab are given an added velocity component so that the centers-of-mass of the slabs move towards each other.  The kinetic energy associated with this initial velocity is $W_{i}=K_{i}=mv_{0}^{2}$. The Nose-Hoover thermostat was applied separately to the two slabs during the dynamical calculation. It was demonstrated by Jarzynksi that the JE can be applied both for isolated systems and thermostatted systems\cite{Jarzynski:1997uj}.

The choice to simulate at $T=500K$ was not made for any specific reason. However, higher temperatures should correspond to increased entropic effects and dissipation, which was an important objective of the present work.  In addition, recent experimental studies of dust aggregation in the initial phases of planetesimal formation have indicated that temperature is an important factor. For example, recent experiments\cite{demirci:2019} with basalt dust grains in the range $T=600$K-$1100$K, exhibit enhanced sticking starting at $T=900$K. Hence, the temperature selected for this article is relevant for the lower temperature regime where dust grains less readily stick.

In the following, lateral translation of the slabs was not included in the initial state, and the center-of-mass velocity imparted to the slabs did not have any lateral component. As a consequence of this choice, the calculations below do not explore the entire phase space which might be expected in an experiment where lateral translation is not so easily controlled. While lateral motion was not constrained in the DCS calculations, it was found that lateral motion (i.e. sliding) of the slabs was minimal. Free energy differences and dissipation computed in this way then should be considered to be relevant only for slabs positioned without including an ensemble of translated slabs. It could be speculated that added translation might reduce the magnitude of $\Delta A$ since the slabs will not always be in ideal registry. On the other hand, because one might expect lateral forces to drive the slabs towards more ideal registry, sliding might occur which could enhance dissipation. Rotation of the slabs was also not included. Subsequent work in this area should more thoroughly consider all different aspects of the available phase space. Inclusion of a more extensive initial phase space will greatly increase the computation required.

\begin{figure}
\begin{center}
\includegraphics[scale=0.25]{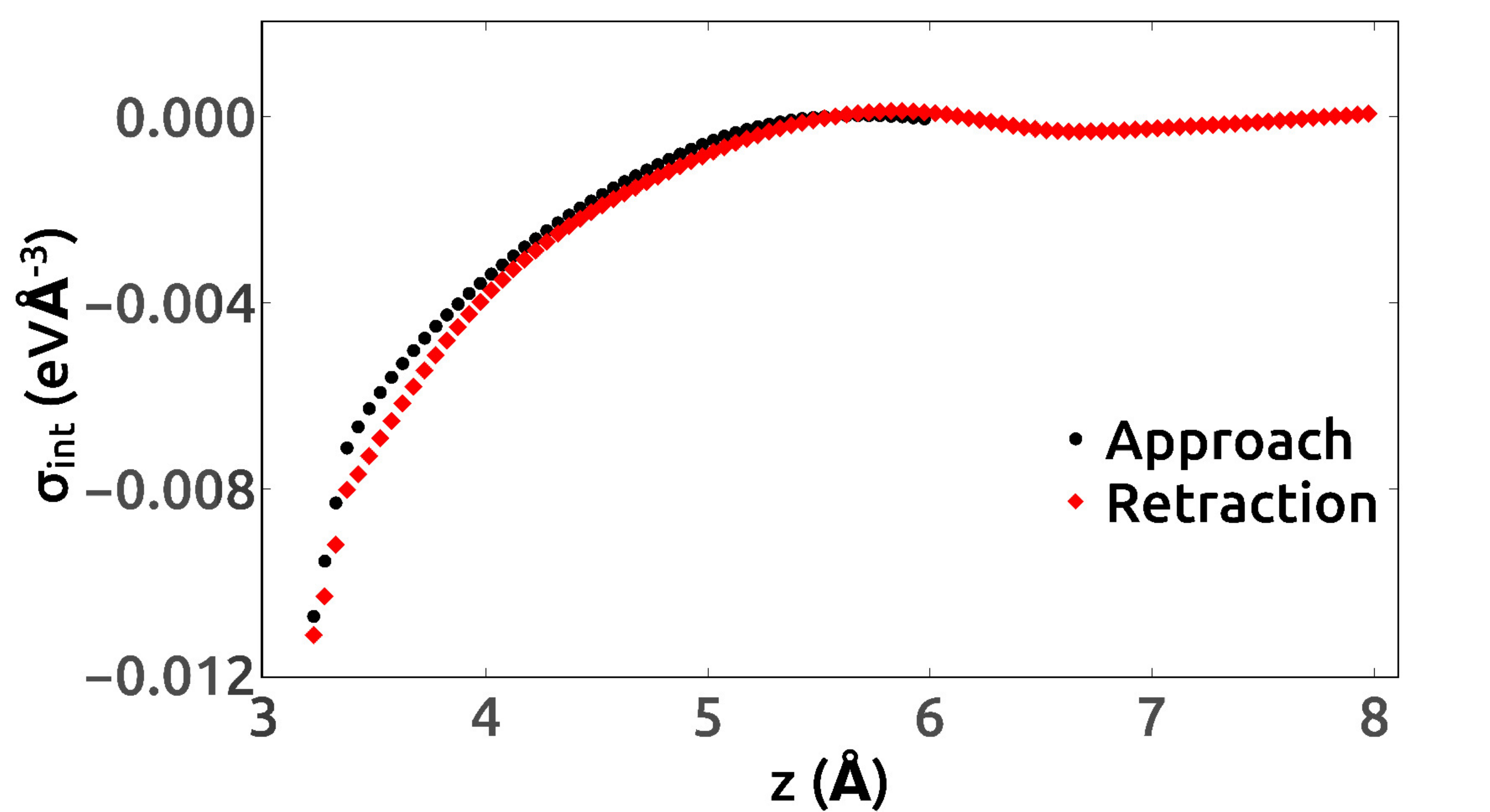}
\end{center}

\caption{
 Hysteresis of the interaction stress $\sigma_{int}=\frac{F_{int}}{S}$ as a function of position $z$ for approach (black circles) and retraction (red diamonds) for $v_{0}=357.5$ms$^{-1}$, $\sigma_{res}=0.0308$eV$\AA^{-3}$, and $S=866.8 \AA^{2}$ with no surface defects.}
 \label{figure2}
\end{figure}

\begin{figure}
\begin{center}
\includegraphics[scale=0.25]{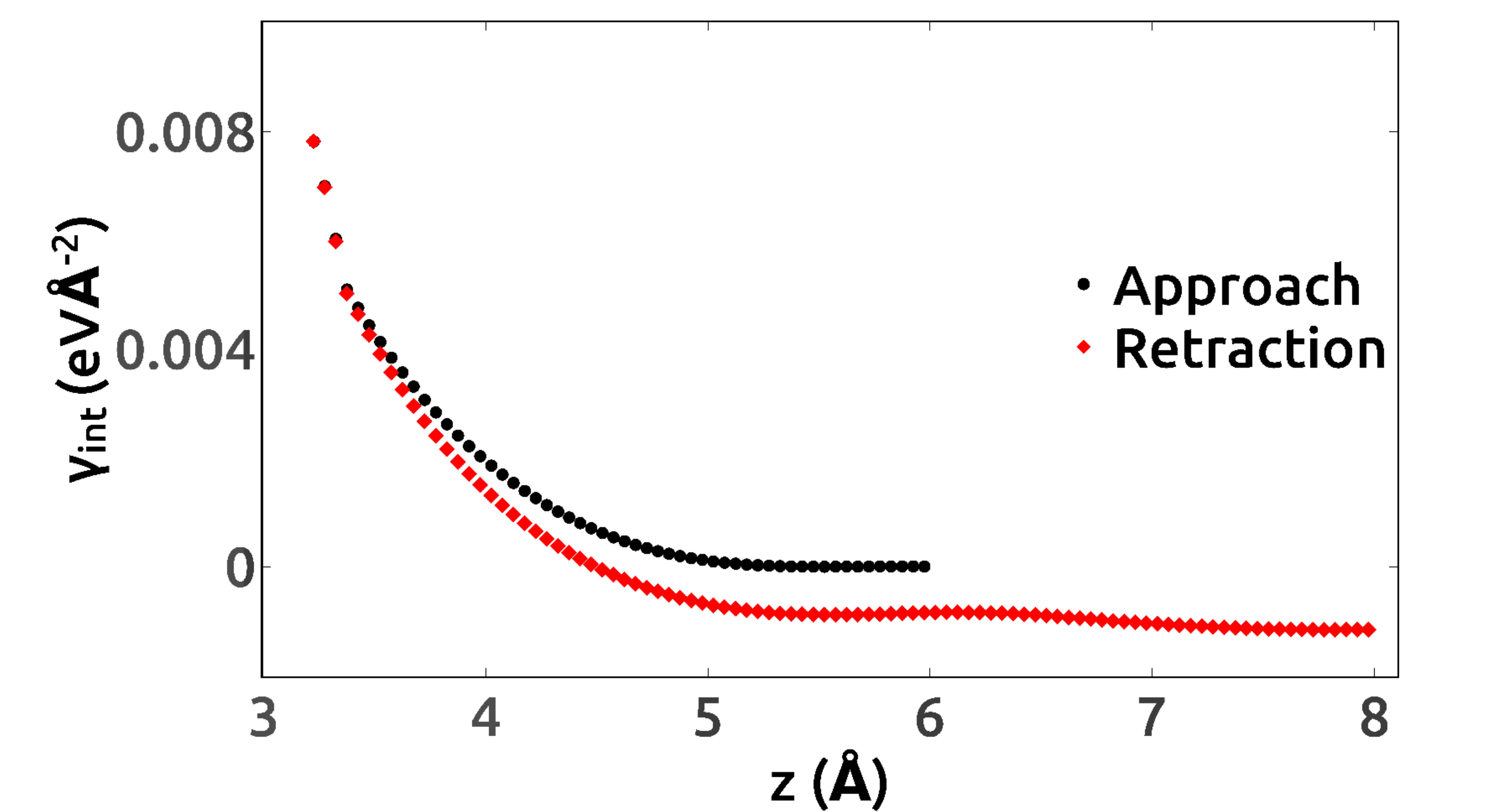}
\end{center}
\caption{
 Interfacial energy $\gamma_{int}=\frac{W_{int}}{S}$ plotted as a function of separation $z$ for approach (black circles) and retraction (red diamonds) for $v_{0}=357.5$ms$^{-1}$, $\sigma_{res}=0.0308$eV$\AA^{-3}$, and $S=866.8 \AA^{2}$ with no surface defects. }
 \label{figure3}
 \end{figure}

In Fig. \ref{figure2}, a representative case of the ensemble-averaged interaction stress $\sigma_{int}(z)=\frac{F_{int}(z}{S}$ is shown as a function of separation $z$ for the approach and retraction phases. The DCS parameters corresponding to results are indicated in the figure caption. Negative values of $\sigma_{int}$ correspond to an attractive interaction between the slabs. The integrated interfacial energy $\gamma_{int}(z)=\frac{W_{int}(z)}{S}$ is obtained as a function of $z$ from the data in Fig. \ref{figure2} by integrating $d \gamma_{int} = \sigma_{int}(z)dz$. The results for $\gamma_{int}(z)$ are shown in Fig. \ref{figure3} for the simulation parameters reported in the caption.
Due to the force hysteresis, $\gamma_{int}(z)$ becomes negative during the retraction phase indicating dissipation. Specifically, the negative values of $\gamma_{int}$ correspond to a loss of the incident kinetic energy.

\begin{figure}
\begin{center}
\includegraphics[scale=0.25]{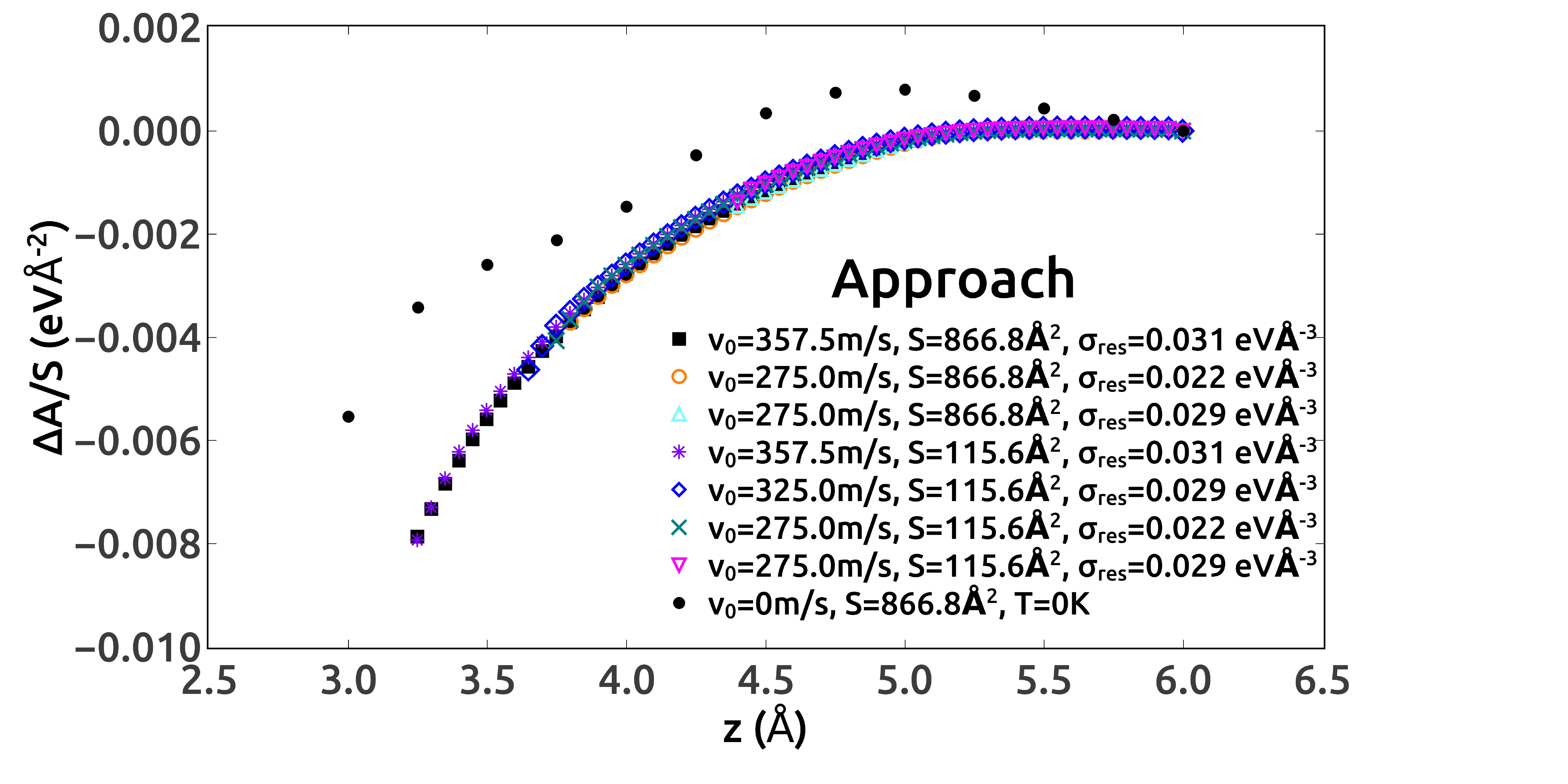}
\end{center}
\caption{
Results for $\frac{\Delta A(z)}{S}$ obtained for the approach phase from the JE (Eq. \ref{JE}). The figure includes each simulated condition (shown in the legend) for slabs without defects. Results for $T=0K$ are also shown.}
\label{figure4}
\end{figure}

\begin{figure}
\begin{center}
\includegraphics[scale=0.25]{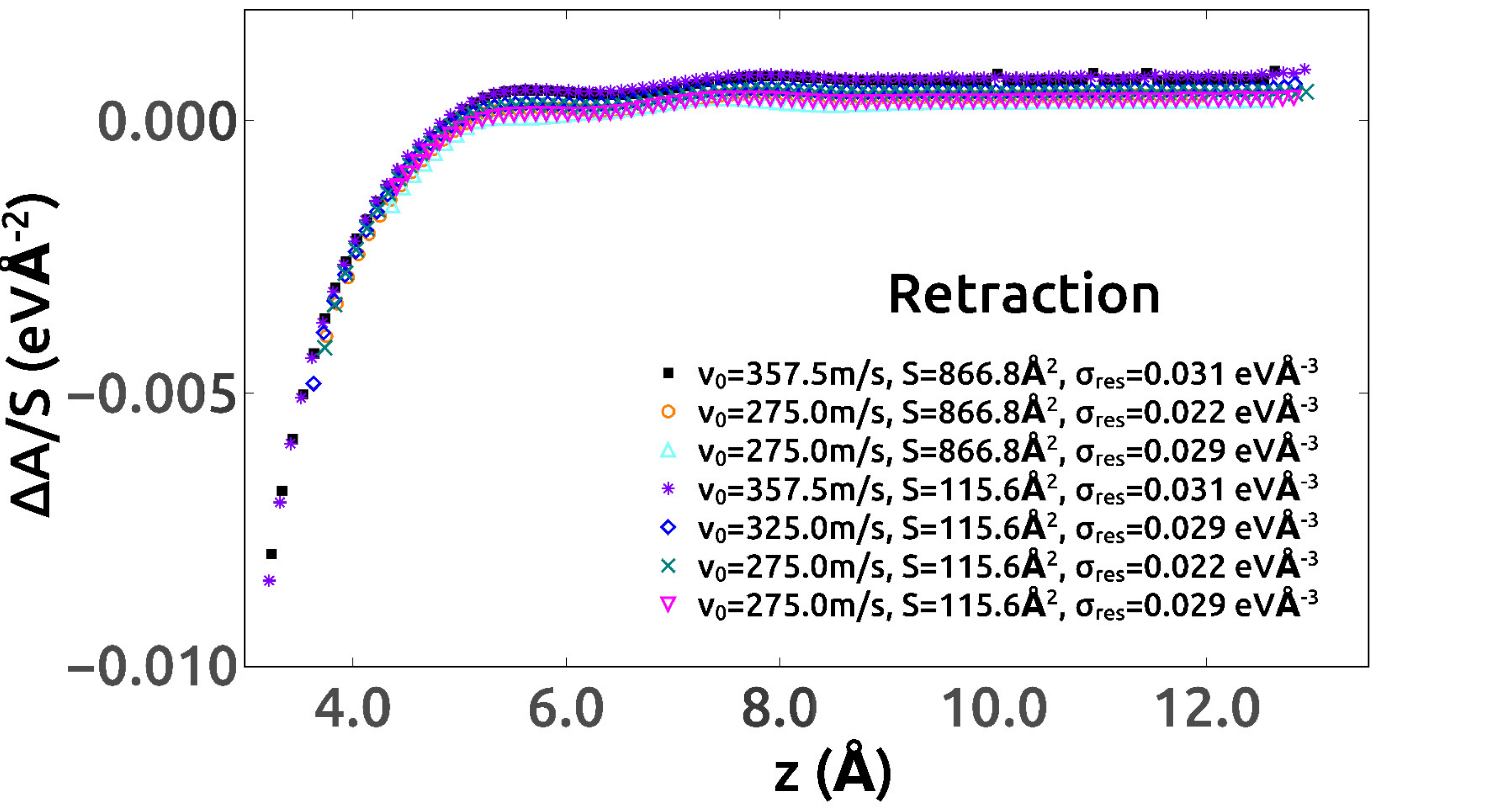}
\end{center}
\caption{
Results for $\frac{\Delta A(z)}{S}$ obtained for the retraction phase from the JE (Eq. \ref{JE}). The figure includes each simulated condition (shown in the legend) for slabs without defects.}
\label{figure5}
\end{figure}

We  next turn to computation of $\Delta A(z) = A(z)-A(z_{0})$, in which $z_{0}$ is the initial separation of the slabs, using the JE in Eq. \ref{JE}. In principle, as long as the structure of the slabs does not vary during the simulation, for example by mass transfer between the slabs, $\Delta A(z)$ should depend only on position $z$ and not on the simulation history. Deviations from this expected behavior could also arise from well-known difficulties encountered in the ensemble average computed in Eq. \ref{JE} \cite{Jarzynski:1997uj,Gore:2003vb}.
The data in Fig. \ref{figure4} and Fig. \ref{figure5} are generated for various both large and small areas $S$, and a range of initial velocities $v_{0}$ and restoring stresses $\sigma_{res}$. In each instance shown in Fig. \ref{figure4} and Fig. \ref{figure5}, the systems were generated without surface MgO defects.  For clarity the approach and retraction phases are shown in Fig. \ref{figure4} and Fig. \ref{figure5} respectively.   Because different system sizes were considered, $\Delta A/S$, where $S$ is the surface area of the slabs, is the quantity shown. The data for the approach phase in Fig. \ref{figure4}  show very close agreement for each value of $v_{0}$, $S$, and $\sigma_{res}$ simulated, suggesting that the JE is likely producing good results for $\Delta A$. The closest approach $z \approx 3.2 \AA$ is achieved for $v_{0}=357.5$ms$^{-1}$ and $\sigma_{res} = 0.0308$eV$\AA^{-3}$ for both surface areas simulated. By contrast, some deviations from completely reversible $\Delta A$ curves are apparent on the retraction phase. In Table \ref{table2} results are shown for $\frac{\Delta A(z)}{S}$ at $z=z_{0}$ during the retraction phase. The results show deviations from the expected result $\frac{\Delta A(z)}{S} =0$ at $z=z_{0}$. The deviations are always positive, and are largest for the large system $S=866.8 \AA^{2}$ and $v_{0}=357.5$ms$^{-1}$. The values of $\Delta A(z)/S$ at $z=z_{0}$ for slabs without defects are shown in Table \ref{table2}.  The deviations for slabs without defects appear to be statistical rather than due to a generation of damage or surface disordering. Specifically, it is known that systems which are further from equilibrium, which in this case should correspond to larger values of $v_{0}$, tend to show greater statistical errors\cite{Jarzynski:1997uj}. This is in part due to the fact that ensemble members with small or even negative dissipation can be rarely sampled in the ensemble yet tend to dominate the averages in Eq. \ref{JE}.

The results in Fig. 4 also include results computed for $T=0K$ obtained by relaxing to zero force for each separation with a constraint force applied evenly to each ion to maintain a fixed relative separation $z$. After relaxing the slabs at a given separation, the resulting structures were moved closer together. In these results there is no entropy changes or dissipation. Comparison of the $T=0K$ results with the DCS calculations at $T=500K$ demonstrate that entropic effects are very important. Because the $T=0K$ calculation is completely reversible, the retraction phase for $T=0K$ is not shown in Fig. 5. The calculation of the interaction between the slabs was continued until a separation $z$ was obtained at which the constraint force was zero. The energy of the system was computed at that point and the interfacial energy $-0.108$eV$\AA^{-2}$ was obtained. If the slabs brought together in this way had made a perfect crystal, this value would be $-2 \gamma_{(010)}=-0.168$eV$\AA^{-2}$. This difference reflects that fact that the surface were rotated with respect to each other as described above and do not make a perfect forsterite crystal when brought together. The $T=0$K interfacial energy provides a reference scale for the  $\frac{\Delta A(z)}{S}$ values throughout the paper.

\begin{table}
\begin{center}
\caption{Results for $\frac{\Delta A(z)}{S}$ at $z=z_{0}$ after retraction for ensembles corresponding to defect-free slabs simulated with different values of $S$, $v_{0}$, and $\frac{F_{res}}{S}$. Values for $\frac{MSE}{S}$ are also given.
}
\begin{tabular} {|c|c|c|c|c|}
\hline
$S$ ($\AA^{2}$) & $v_{0}$ (ms$^{-1}$) & $\sigma_{res}$ (eV $\AA^{-3}$) & $\frac{\Delta A}{S}\times 10^{4}$ (eV $\AA^{-2}$) &
$\frac{MSE}{S} \times 10^{4}$ (eV $\AA^{-2}$)  \\\hline 
115.6 &  275.0 & 0.0221  &  2.60 & 0.20  \\\hline
115.6 &  275.0   &0.0291  &  1.10   & 0.31\\\hline
115.6 &  325.0   & 0.0291  &  3.00  & 0.30 \\\hline
866.8 & 275.0   & 0.0221   & 2.20 & 0.16 \\\hline
866.8 & 275.0   & 0.0291   & 0.99  & 0.04 \\\hline
866.8 & 357.5   & 0.0308   & 5.10  & 1.07 \\\hline
\end{tabular}
\label{table2}
\end{center}
\end{table}

\begin{figure}
\begin{center}
\includegraphics[scale=0.25]{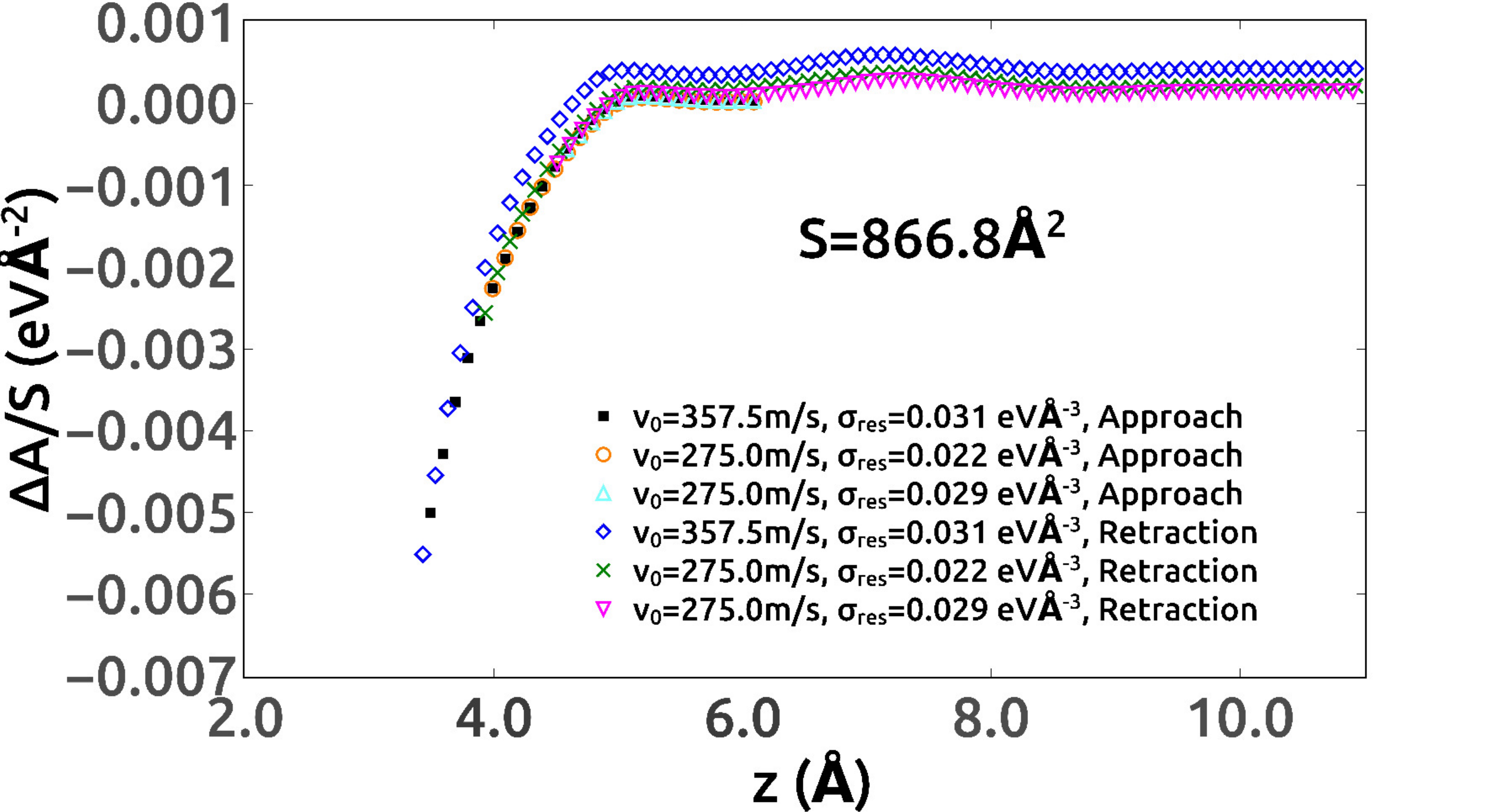}
\end{center}
\caption{
Results for $\frac{\Delta A}{S}$ as a function of separation $z$ obtained for large slabs $S=866.8\AA^{2}$ with MgO defects. 
Data compiled together for ensembles with different values of $\sigma_{res}$ and $v_{0}$ as indicated in the legend. }
\label{figure6}
\end{figure}

\begin{figure}
\begin{center}
\includegraphics[scale=0.25]{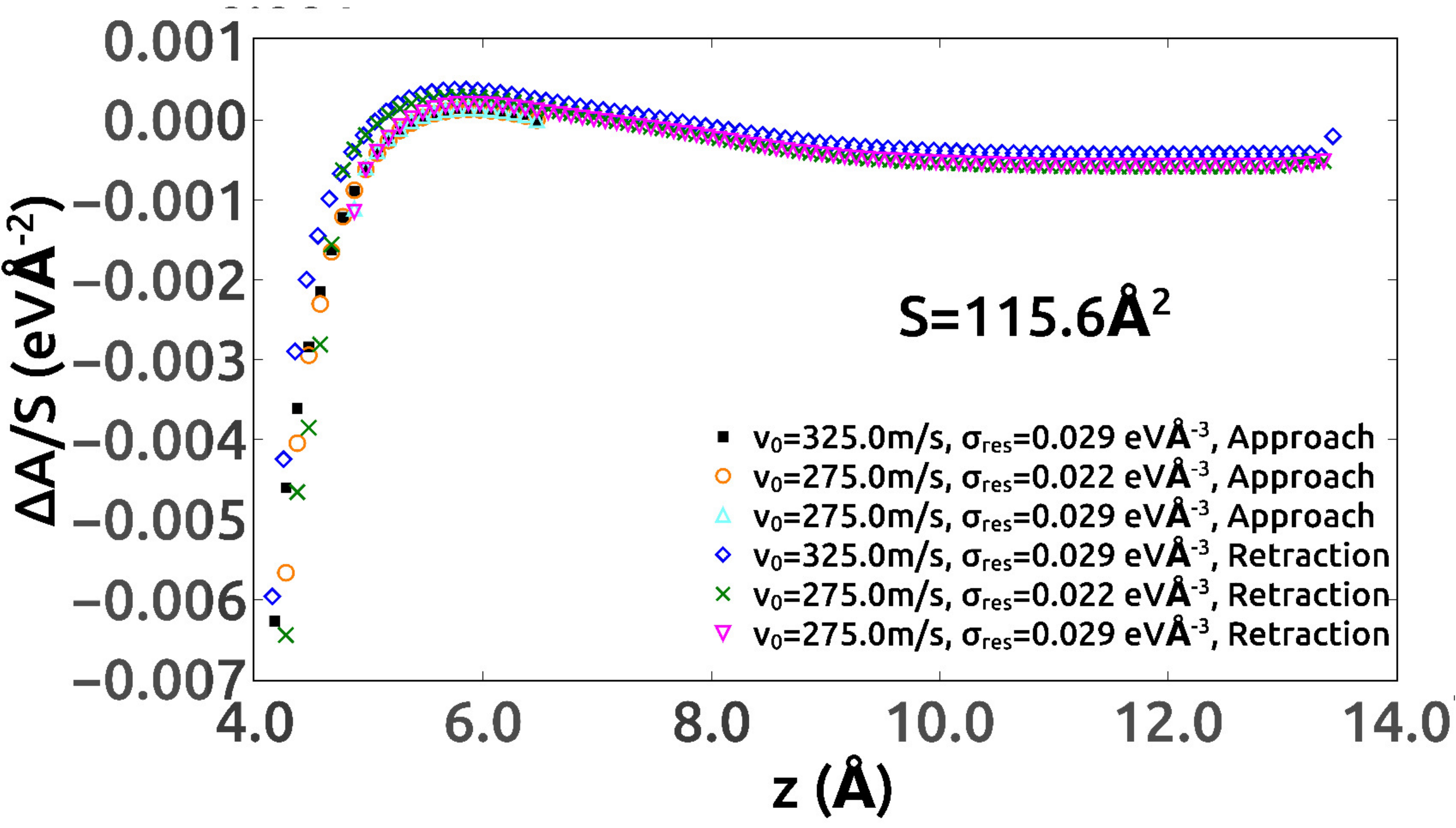}
\end{center}
\caption{
Results for $\frac{\Delta A}{S}$ as a function of separation $z$ obtained for small area slabs with $S=115.6\AA^{2}$ with MgO-vacancy defects. 
Data compiled together for ensembles with different values of $\sigma_{res}$ and $v_{0}$ as indicated in the legend.
Significant deviations for reversibility are seen for $v_{0}=325$ms$^{-1}$.}
\label{figure7}
\end{figure}

Plots for $\frac{\Delta A(z)}{S}$ corresponding to systems with MgO-vacancy defects are shown in Fig. \ref{figure6} for $S=866.8 \AA^{2}$ and Fig. \ref{figure7}  for $S=115.6 \AA^{2}$. In these figures, both the approach and retraction phases are shown together for comparison. An additional distinction between the two slab areas is that the small $S=115.6\AA^{2}$ contained a higher density of MgO defects in comparison to slabs with area $S=866.8 \AA^{2}$. Consequently,  simulation results using the two different slab areas were not expected to generate the same free energy curve $\Delta A(z)$. In fact, comparison of Fig. \ref{figure6}  and Fig. \ref{figure7}  show significant differences. One important difference observed is that the small slabs with area $S=115.6\AA^{2}$, plotted in Fig. \ref{figure7} , have much more negative values of ${\Delta A \over S}$ when compared to the data shown in Fig. \ref{figure6} . This indicates that the higher density of surface defects in the small slabs tend to result in stronger adhesive interactions between the slabs. While this is true for the reversible part of the interaction described by $\Delta A(z)$, as we will see, the dissipation is also stronger in the presence of a higher density of defects.

For large $S=866.8 \AA^{2}$ interfaces with MgO-vacancy defects, generally $\Delta A(z)$ curves agreed reasonably well for approach and retraction phases for different DCS parameters $v_{0}$ and $\sigma_{res}$. The system with $v_{0}=357.5$ms$^{-1}$ and $\sigma_{res}=0.031$eV$\AA^{-3}$ showed the least reversible $\Delta A(z)$ graph. This is quite similar to the plots in Fig. \ref{figure4} and Fig. \ref{figure5} for slabs without defects. Namely, the system with DCS parameters that involve the largest speed $v_{0}$ and closest distance of approach corresponds to the largest deviation from complete reversibility. Results for $\Delta A(z)$ at $z=z_{0}$ during the retraction phase are shown in Table \ref{table3}. Again, values are all systematically positive and larger than the expected result $\Delta A=0$. 

One striking difference in the small slabs with MgO-vacancy defects is the substantial negative values $\frac{\Delta A(z)}{S}\sim -5 \times 10^{-4}$ eV $\AA^{-2}$ for large separations $z>z_{0}$. This is evident from Fig. \ref{figure7}. In contrast to the tendency for systematic bias $\Delta A>0$ for $z=z_{0}$ and beyond, this suggests qualitatively different behavior. In fact, what we observe is that systems can disorder and find lower energy minima after retraction. Therefore, the structures obtained after retraction exhibit somewhat different bonding arrangements and $\Delta A <0$ occurs because the structures have a lower potential energy. Generally, we find that surface structures even before the approach phase can have different potential energy values, and that the structures explore many different local minima during each calculation. The presence of different local-energy minima and the ability of the system to explore those minima was established by annealing structures to $T=0K$ both before the approach phase and after retraction. This observation of multiple local minima indicates that the ensemble represents a quasi-equilibrium distribution, which is discussed in context of the JE in previous applications to free energy differences in glasses\cite{Williams_2008}. In more extreme cases, 
 to be detailed later on, mass transfer between the slabs is observed. Specifically, mass transfer is found to occasionally occur between the smaller slab simulations with area $S=115.6AA^{2}$ and a higher concentration of MgO-vacancy defects. As we will see, these observations are also associated with stronger dissipation and long-tailed kinetic energy distributions. In short, many of the simulations with MgO-vacancy defects are very strongly dissipative and are found to be in a far-from-equilibrium state.

The Jarzynksi estimator defined in Eq. \ref{JE} is known to be susceptible to biasing due to nonlinear averaging. The convergence properties associated with the expression in Eq. \ref{JE} was extensively explored  by Gore et al \cite{Gore:2003vb}. This was done by accurately computing $\Delta A$ for a very large ensemble of trajectories, and then a systematic evaluation of the error, including biasing, by evaluating the statistical distribution of estimates for $\Delta A(N)$ using a much smaller number of trajectories. It was demonstrated\cite{Gore:2003vb} that Eq. \ref{JE} systematically biases $\Delta A(N)$ to more positive values in comparison to the exact value $\Delta A$ obtained for a very large number of trajectories. One can easily show that for $N=1$, the bias is equal to the average dissipative work $W_{diss}$\cite{Gore:2003vb}. As $N$ increases, the bias decreases, but in some cases very slowly. In addition to biased values, as with any ensemble property, $\Delta A(N)$ also has an associated variance. The mean-squared error $MSE(N)$ for ensembles with $N$ trajectories was defined previously \cite{Gore:2003vb} using,
\begin{equation}
MSE(N) = \langle \left(\Delta A(N) - \Delta A \right)^{2} \rangle = \sigma^{2}(N) + B^{2}(N)
\label{MSE}
\end{equation}
in which $\sigma^{2}(N)$ represents the variance and $B^{2}(N)$ the bias associated with ensembles with $N$ trajectories. The bias can be understood to be the systematic deviation of the free energy $B(N)=\langle \Delta A(N) \rangle - \Delta A$. In these expressions, $\Delta A$ is the accurate value of the free-energy difference. The variance is given by $\sigma_{N}= \langle (\Delta A(N) - \langle \Delta A(N) \rangle)^{2} \rangle$. The free energy estimates are made from $M$ trials, each with $N$ members in the ensemble. To evaluate Eq. \ref{MSE}, a large number of trials $M$ is required. The specific calculations required are given by,
\begin{equation}
\langle \Delta A(N) \rangle = \lim_{M \rightarrow \infty} \frac{1}{M} \sum_{k=1}^{M} \Delta A_{k}(N)
\label{DA}
\end{equation}
\begin{equation}
\langle \Delta A^{2}(N) \rangle = \lim_{M \rightarrow \infty} \frac{1}{M} \sum_{k=1}^{M} \Delta A_{k}^{2}(N)
\label{DA2}
\end{equation}
in which $\Delta A_{k}(N)$ represents the estimate from trial $k$ for $\Delta A$ computed using Eq. \ref{JE} with $N$ ensemble members.
The systematic behavior of both the bias and variance terms was explored previously for near equilibrium trajectories and Gaussian work distributions\cite{Gore:2003vb}. In the present case, as will be demonstrated, the work distributions vary strongly from Gaussians, and hence some of the analysis methods \cite{Gore:2003vb} may be of limited use. Nevertheless, it is worth noting that Gore et al\cite{Gore:2003vb} demonstrated that bias values converge with $N$ as $B(N) \approx \frac{W_{diss}}{N^{\alpha}}$. The exponent $\alpha$ was shown to strongly depend on $W_{diss}$, and could become significantly smaller than $1$. For example, it was found previously \cite{Gore:2003vb} that $\alpha \sim 0.15$ for $W_{diss} \approx 15k_{B}T$, which corresponds to typical values in our calculations.

The basic idea that we use, which is consistent with the previous analysis\cite{Gore:2003vb}, is to explore the statistical properties obtained from several independent ensembles which are subsets of the entire ensemble of trajectories. Consistent with the description above, we refer to each of these subsets as ``trials''. In particular, for the small systems, which have data for 500 trajectories, we explored the statistics of $M=25$ trials with $N=20$ ensemble members in each trial. For the larger systems, which have data for 200 trajectories, we explored the statistics of $M=10$ trials with $N=20$ ensemble members in each trial. The analysis of $MSE(N)$ then uses Eqs. \ref{MSE}, \ref{DA}, \ref{DA2} but with a rather small number of trials $M$. The value of $\Delta A$ used in Eq. \ref{MSE} is the best result obtained from the entire ensemble. Hence, $MSE(N)$ represents the statistical error associated with each trial. Since the results we present in this paper are actually averages over the entire ensemble, or equivalently over all the trials, we should expect that the error in our results are smaller than the $MSE(N)$ estimates. 

In Table \ref{table2} and Table \ref{table3}, we report values for $MSE(N)$ defined in Eq. \ref{MSE}. Because the results averaged over all trajectories show clear evidence of bias, the $MSE(N)$ values predominantly are due to variance rather than bias. The bias is already reflected in deviation from $\Delta A=0$ for the closed cycle. It is seen that the $MSE(N)$ is substantially less than the bias. Therefore, bias tends to dominate over variance. Finally, it has been shown that bias tends to result in underestimated values for $W_{diss}$ \cite{Gore:2003vb}. This is apparent from Eq. \ref{wdissnotclosed}. Specifically, Eq. \ref{wdissnotclosed} shows that when $\Delta A(z)$ is systematically biased towards positive values, then $W_{diss}(z)$, and hence $\gamma_{diss}(z)$ curves, tend to underestimate dissipation. To reduce bias, it is quite possible that the number of trajectories in the ensemble would need to greatly increase. 

In summary, the statistical analysis above corresponds primarily to an estimate for $\sigma^{2}(N)$ rather than the bias term $B^{2}(N)$ in Eq. \ref{MSE}. The analysis shows that variance $\sigma^{2}(N)$  is quite small, and the variance for the entire ensemble of trajectories must be even smaller. However, other measures, primarily the deviation from $\Delta A=0$ for a closed cycle, indicate that the bias term $B^{2}(N)$ important. The analysis above depends on the already biased values of $\Delta A$, and hence is not able itself to determine $B^{2}(N)$.

\begin{table}
\begin{center}
\caption{Results for $\frac{\Delta A}{S}$ at $z=z_{0}$ after retraction for ensembles corresponding to  slabs with MgO-vacancy defects simulated with different values of $S$, $v_{0}$, and $\sigma_{res}=\frac{F_{res}}{S}$. Values for $\frac{MSE}{S}$ are also given.
}
\begin{tabular} {|c|c|c|c|c|}
\hline
$S$ ($\AA^{2}$) & $v_{0}$ (ms$^{-1}$) & $\sigma_{res}$ (eV $\AA^{-3}$) & $\frac{\Delta A}{S} \times 10^{4}$ (eV $\AA^{-2}$) &
$\frac{MSE}{S} \times 10^{4}$ (eV $\AA^{-2}$) \\\hline 
115.6 &  275.0 &  0.0221  &  1.66  &   1.18 \\\hline
115.6 &  275.0 &  0.0285  &  1.24 &  0.31 \\\hline
115.6 &  325.0 &  0.0285  & 2.62 &   1.39 \\\hline
866.8 & 275.0 &  0.0220   & 1.60 &  0.28 \\\hline
866.8 & 275.0 &  0.0287   & 0.56 &  0.05 \\\hline
866.8 & 357.5 & 0.0307   & 3.32  & 1.06 \\\hline
\end{tabular}
\label{table3}
\end{center}
\end{table}

\begin{figure}
\begin{center}
\includegraphics[scale=0.25]{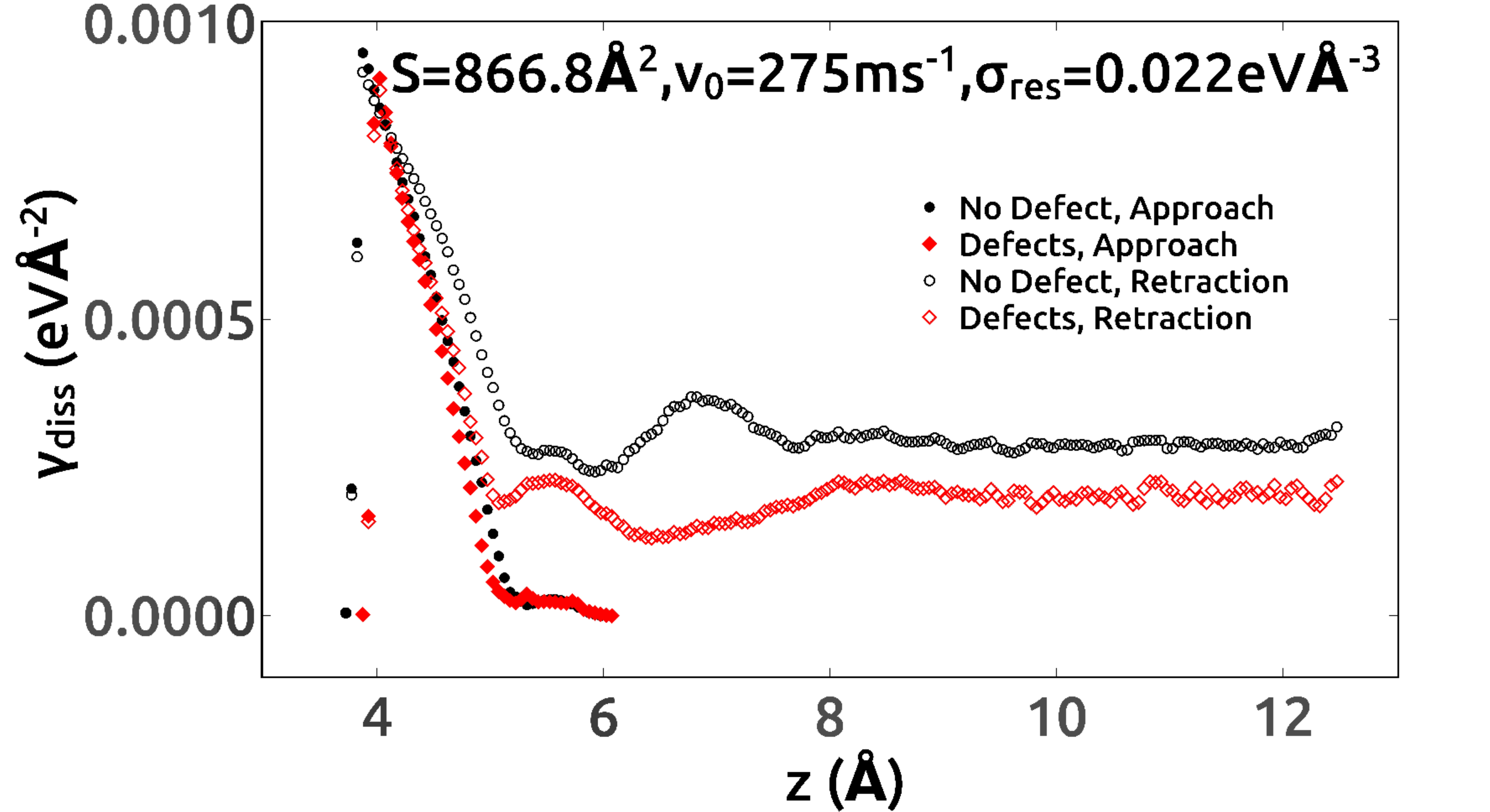}
\end{center}
\caption{
Dissipation $\gamma_{diss}$ obtained from the JE as a function separation $z$ for defect free slabs (filled black circles and open black circles for approach and retraction respectively) and slabs with MgO defects (filled red diamonds and open red diamonds for approach and retraction respectively). Simulation conditions shown in the figure.}
\label{figure8}
\end{figure}

Dissipation can be obtained  from the JE as written in Eq. \ref{wdissnotclosed} as a function of separation $z$. This calculation requires relatively accurate values of $\Delta A(z)$.
Calculation of $\gamma_{diss}(z)=\frac{W_{diss}(z)}{S}$ was performed for different DCS conditions as a function of slab separation $z$. In Fig. \ref{figure8}, results for large slabs $S=866.8\AA^{2}$ are shown. To elucidate the role played by MgO-vacancy defects, Fig. \ref{figure8} compares $\gamma_{diss}(z)$  curves for systems both with and without surface MgO-vacancy defects but otherwise with identical DCS parameters. Several consistent trends emerge. First, the approach phase shows a large increase in $\gamma_{diss}(z)$ as separation $z$ decreases. During the approach, $\gamma_{diss}(z)$ does not depend strongly on the presence of defects. 
The retraction phase shows interesting behavior. Specifically, $\gamma_{diss}(z)$ initially decreases as the slab separation $z$ increases. This surprising result is analyzed further below. 
 Finally, $\gamma_{diss}(z)$ shows a dependence on MgO-vacancy defects during the retraction phase, with $\gamma_{diss}(z)$ smaller when MgO-vacancy defects are present. This is likely due to the fact that for perfect slabs, the interaction energy between the slabs is larger, since a regular interface could be formed if $z$ were allowed to decrease to even smaller values. 

\begin{figure}
\begin{center}
\includegraphics[scale=0.25]{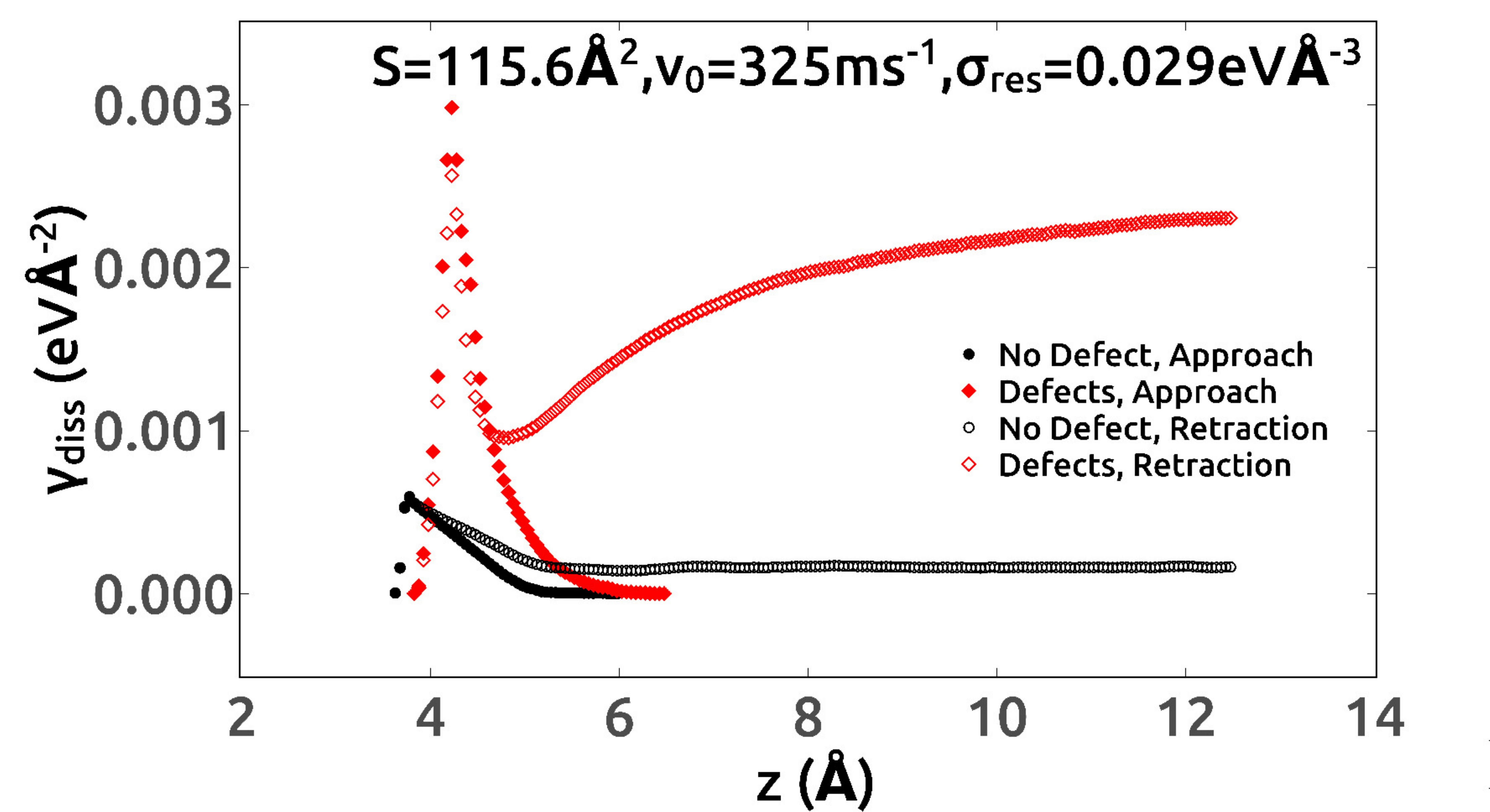}
\end{center}
\caption{
Dissipation $\gamma_{diss}$ obtained from the JE as a function separation $z$ for defect free slabs (filled black circles and open black circles for approach and retraction respectively) and slabs with MgO-vacancy defects (filled red diamonds and open red diamonds for approach and retraction respectively). Simulation conditions shown in the figure.}
\label{figure9}
\end{figure}

However, in contrast to the above results, the smaller slabs with area $S=115.6\AA^{2}$ that contain a higher density of surface MgO vacancies behave very differently.
The indication of strikingly different behavior was already indicated by comparison of the $\Delta A/S$ plots in Fig. \ref{figure6} and Fig. \ref{figure7}, which indicate larger reversible adhesion behavior when high defect densities are present. The large dissipation seen for small slabs with defects can be seen in Fig. \ref{figure9}, where the results from calculations both with and without defects are shown. Some trends in how $\gamma_{diss}(z)$ depends on $z$ are qualitatively the same as those seen in Fig. \ref{figure8}, but the very slow increase in dissipation for $z$ beyond $5\AA$ during retraction is a distinct feature associated with higher defect densities. It will be demonstrated that this behavior is due to the tendency of ions in near vacant sites to be more strongly displaced by the interaction process, including the possibility of ion exchange between the two slabs. In contrast to the defective surfaces, slabs with perfectly crystalline surfaces do not show large structural disruptions or ion exchange for the DCS parameters used.

Considerably more accurate calculations of dissipation can be obtained from force hysteresis curves, specifically by integrating to find the interaction work associated with a closed cycle of approach and retraction. Values of  $-\gamma_{int}$ obtained during the retraction phase at $z=z_{0}$ are shown in Table \ref{table4} and Table \ref{table5} for systems with and without MgO-vacancy defects respectively. As described previously, $-\gamma_{int}$ is obtained directly from integration of the force hysteresis curves, and represents the most accurate results for the dissipation. However, the dissipation is only obtained at $z=z_{0}$ since a closed cycle is required. In addition, the data in Tables \ref{table4}-\ref{table5} show the initial kinetic energy divided by the surface area $K_{i}/S$. This quantity is useful for establishing the fraction of incident energy that is dissipated during approach and retraction phases.

The largest values of $-\gamma_{int}$ are seen to occur for small systems with MgO-vacancy defects. This appears to be due to the fact that high defect densities result in more weakly-bound surface ions that are easily displaced from their lattice sites. In some instances, analyzed more thoroughly below, transfer of matter between the slabs is found to occur. Finally, the interfacial dissipation quantified by $-\gamma_{int}$ can be compared to the kinetic energy normalized by surface area, $K_{i}/S$. Specifically, the strongest dissipation is reported in Table \ref{table5} involves the dissipation of $\sim 2.8 \%$ of the incident kinetic energy. Interestingly, in this case the surfaces do not approach closer than $z \sim 4\AA$.  By contrast, simulations of slabs without MgO-vacancy defects can approach to $z \sim 3.2 \AA$, but involve dissipation at most $\sim 1\%$ if the incident kinetic energy $K_{i}$. This indicates an important role for surface defects in the dissipation process. In fact, even stronger dissipation occurs if surfaces are allowed to approach more closely. Specifically, later in the article we describe simulations with DCS parameters that result in some cases of sticking which are not reported in Table \ref{table4} or Table \ref{table5}. 

\begin{table}
\begin{center}
\caption{Results for $-\gamma_{int}$  at separation $z=z_{0}$ after retraction for ensembles corresponding to  slabs without defects simulated with different values of $S$, $v_{0}$, and $\sigma_{res}=\frac{F_{res}}{S}$. Also shown is the quantity $\frac{K_{i}}{S}$ which is the kinetic energy of the intial translation motion per unit area.
}
\begin{tabular} {|c|c|c|c|c|}
\hline
$S$ ($\AA^{2}$) & $v_{0}$  (ms$^{-1}$)  & $\frac{K_{i}}{S}\times 10^{2}$(eV$\AA^{-2}$)&  $\sigma_{res}$ (eV $\AA^{-3}$)  & $-\gamma_{int}  \times 10^{4}$ (eV$\AA^{-2}$) \\\hline 
115.6 & 275.0   & 4.579  & 0.0221   & 2.67 \\\hline
115.6 & 275.0  & 4.579  & 0.0291     & 0.57 \\\hline
115.6 & 325.0  & 6.396  & 0.0291      & 3.05 \\\hline
866.8 & 275.0  & 4.581  & 0.0221     & 3.71\\\hline
866.8 & 275.0  & 4.581  & 0.0291     & 0.82 \\\hline
866.8 & 357.5  & 7.741  & 0.0308    & 8.53 \\\hline
\end{tabular}
\label{table4}
\end{center}
\end{table}

\begin{table}
\begin{center}
\caption{Results for $-\gamma_{int} $  at separation $z=z_{0}$ after retraction for ensembles corresponding to  slabs with MgO-vacancy defects simulated with different values of $S$, $v_{0}$, and $\sigma_{res}=\frac{F_{res}}{S}$. Also shown is the quantity $\frac{K_{i}}{S}$ which is the kinetic energy of the initial translation motion per unit area.
}
\begin{tabular} {|c|c|c|c|c|}
\hline
$S$ ($\AA^{2}$) &$v_{0}$  (ms$^{-1}$) &  $\frac{K_{i}}{S}\times 10^{2}$(eV$\AA^{-2}$)&  $\sigma_{res}$ (eV $\AA^{-3}$)  & $-\gamma_{int}  \times 10^{4}$ (eV$\AA^{-2}$) \\\hline
115.6 & 275.0 & 4.497  & 0.0221    & 5.28 \\\hline
115.6  & 275.0  &4.497  & 0.0285     & 1.15 \\\hline
115.6  & 325.0 &6.279  & 0.0285   & 17.38 \\\hline
866.8  & 275.0 &4.561  & 0.0220     & 2.70\\\hline
866.8  & 275.0 &4.561  & 0.0287  & 0.45\\\hline
866.8  & 357.5 &7.708  & 0.0307    & 6.68 \\\hline
\end{tabular}
\label{table5}
\end{center}
\end{table}
 
\begin{figure}
 \begin{center}
\includegraphics[scale=0.25]{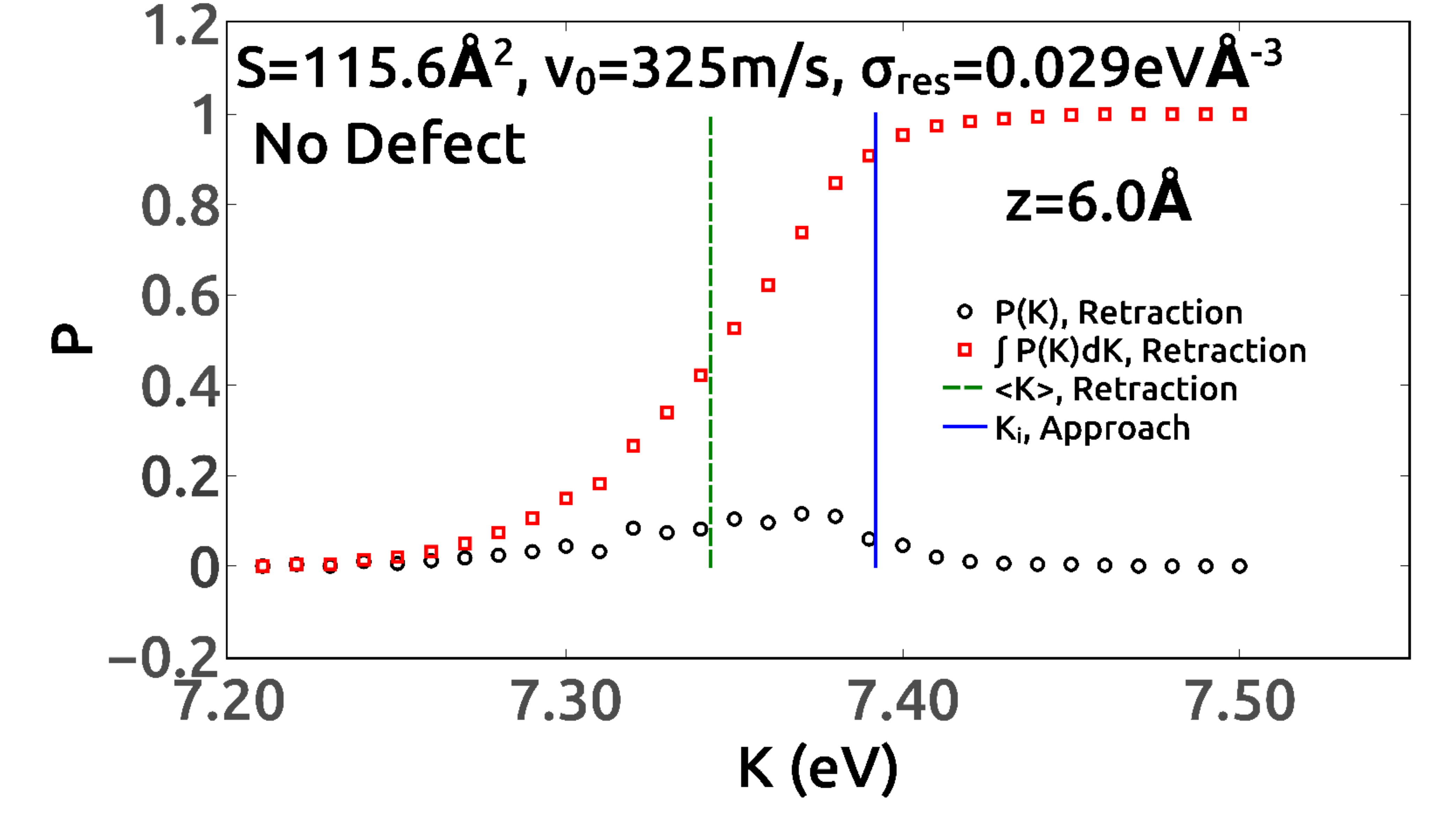}
\end{center}
\caption{
Kinetic energy distribution $P(K)$ shown for $z=6.0\AA$ (open black circles)  during the retraction phase for a small slab $S=115.6 \AA^{2}$ without defects. The running integrals $\int P(K) dK$ is also shown (open red squares). The average $\langle K \rangle$ for retraction (dashed green line) is shown. The value $K_{i}$ is also shown (solid blue line).  }
\label{figure10}
\end{figure}

 Additional information is obtained by analysis of the kinetic energy distribution function for various points along a trajectory. First we focus on systems without defects where dissipation is relatively small. In Fig. \ref{figure10}, the kinetic energy distribution $P(K)$ is shown for $S=115.6 \AA^{2}$ and $v_{0}=325$ms$^{-1}$ and $\sigma_{res}= 0.291$eV$\AA^{2}$. Results for the distribution $P(K)$ are shown during retraction at $z=z_{0}$, along with the initial kinetic energy $K_{i}$. The distribution itself does not appear to be Gaussian, with a few members with rather large dissipation. The  average value $\langle K \rangle$ is also shown, from which the dissipation for the closed cycle can be obtained. The dissipation is obtained from the difference $K_{i}-\langle K \rangle \sim 0.04$eV. This corresponds to the value $-\gamma_{int}$ shown in Table 4. It can be seen that very few systems have zero or negative dissipative work. This result is rather consistent with other simulations without defects. 
 
\begin{figure}
 \begin{center}
\includegraphics[scale=0.25]{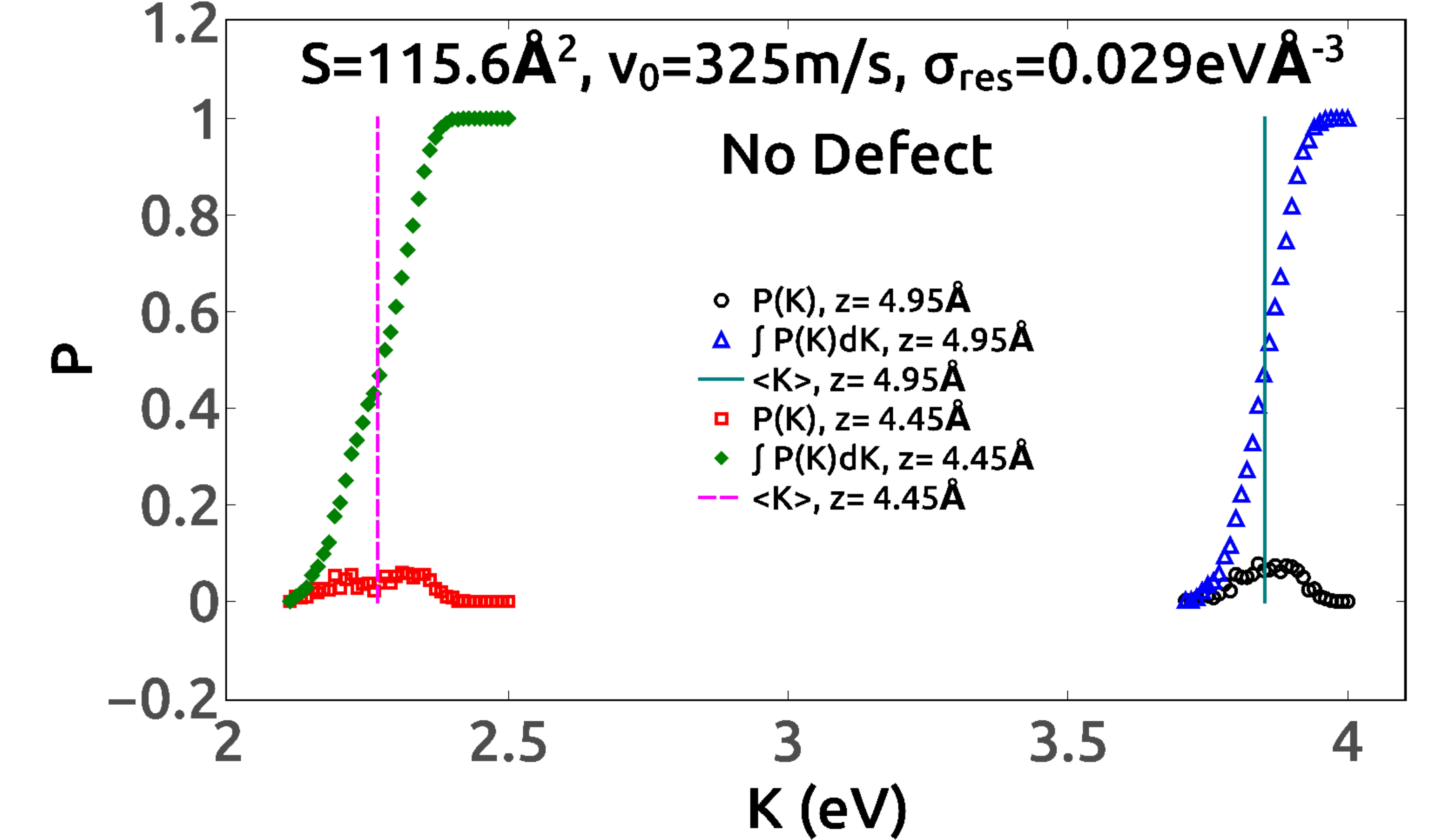}
\end{center}
\caption{
Kinetic energy distribution $P(K)$ shown for two different $z$ coordinates during the retraction phase for small slabs $S=115.6 \AA^{2}$ without defects. Results for  $z=4.95\AA$ (open black circles) and $z=4.45\AA$ (open red squares) show a narrowing distribution as $z$ increases.  Running integrals $\int P(K) dK$ are shown for the two cases (open blue triangles for $Z=4.95\AA$ and closed green diamonds for $z=4.45\AA$). Average kinetic values $\langle K \rangle$ for $z=4.95\AA$ (dashed magneta line) and $z=4.45\AA$ (solid blue line) are shown.  Simulation parameters correspond to 
$v_{0}=325$ms$^{-1}$ and $\sigma_{res}=0.029$eV$\AA^{-3}$. }
\label{figure11}
\end{figure}

 We next explore the issue of the decreasing values of $\gamma_{diss}(z)$  with increasing $z$ during the retraction phase. This behavior was seen in all cases. Based on the JE analysis, decreasing values of $\gamma_{diss}$ must correspond to decreasing width of the kinetic energy distribution. Specifically, although the distributions are not strictly Gaussian, for cumulant expansions of the JE the second order term is proportional the the variance\cite{Jarzynski:1997uj,Gore:2003vb}. In Fig. \ref{figure11}, the kinetic energy distributions are plotted for two different separations during the retraction phase. The values of $z$ were chosen to correspond to the portion of the retraction phase where $W_{diss}$ is decreasing with increasing separation. The results clearly show that the distribution of kinetic energy values narrows as separation $z$ increases, consistent with the decrease in $\gamma_{diss}(z)$ with increasing $z$ shown by the JE. In the calculations reported here, while the distributions deviate significantly from Gaussian distributions, the same qualitative connection between the width of the distribution and $\gamma_{diss}(z)$ should apply. In the discussion section, we will develop an argument which connects this surprising behavior to stick-slip motion generally encountered in the context of sliding friction\cite{Socoliuc:2004wx}.
 
\begin{figure}
\begin{center}
\includegraphics[scale=0.25]{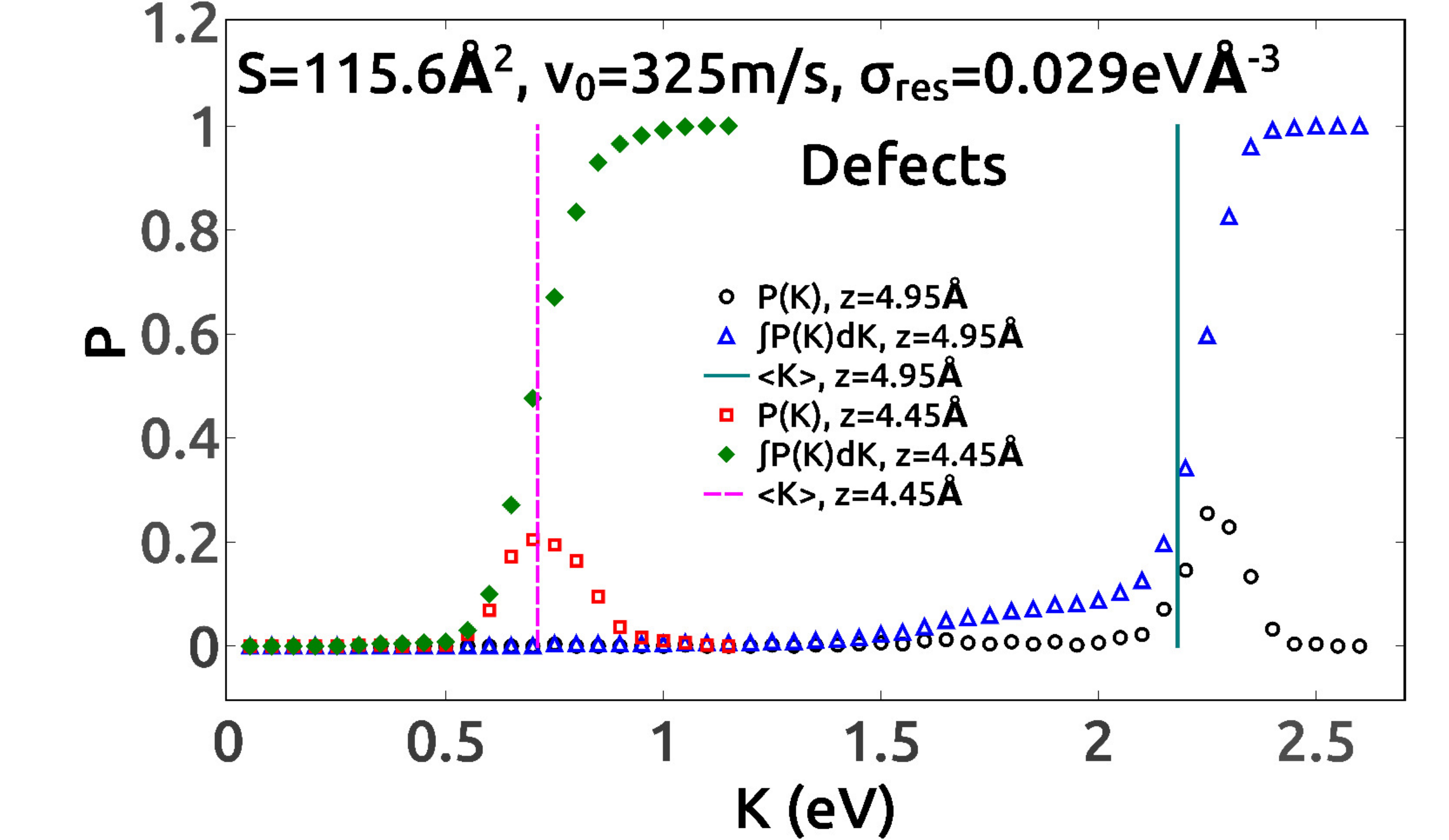}
\end{center}
\caption{
Kinetic energy distribution $P(K)$ shown for $z=4.95\AA$ (open black circles) and $z=4.45\AA$ (open red squares) during the retraction phase for a small slab $S=115.6 \AA^{2}$ with MgO-vacancy defects. Running integrals $\int P(K) dK$ are also shown for $z=4.95\AA$ (open blue triangles) and $z=4.45 \AA$ (filled green diamonds).  As the system pulls further away during retraction, a tail in the distribution develops at lower values of $K$.}
\label{figure12}
\end{figure}

When MgO-vacancy defects are present, some aspects of the qualitative behavior change, and as the data in Table \ref{table5} and Fig. \ref{figure9} demonstrate, dissipation can become substantially more significant. First, the slabs experience dissipation at separations up to $\sim 10\AA$, as shown previously in Fig. \ref{figure9}. The mechanism for this behavior is found by examination of kinetic energy distributions with increasing separation, as shown in Fig. \ref{figure12}. As the separation $z$ increases, some systems in the ensemble (with probability $\sim 0.20$) exhibit substantially larger dissipation. This shows here in the long tail in the distribution $P(K)$ at lower kinetic energy values, which is most easily seen by plotting the running integral $\int P(K)dK$. The mean of the distribution at $z=4.95\AA$ in Fig. 12 is substantially displaced from the median of the distribution. This behavior indicates stronger dissipation and an ensemble much further from equilibrium. 

\begin{figure}
\begin{center}
\includegraphics[scale=0.25]{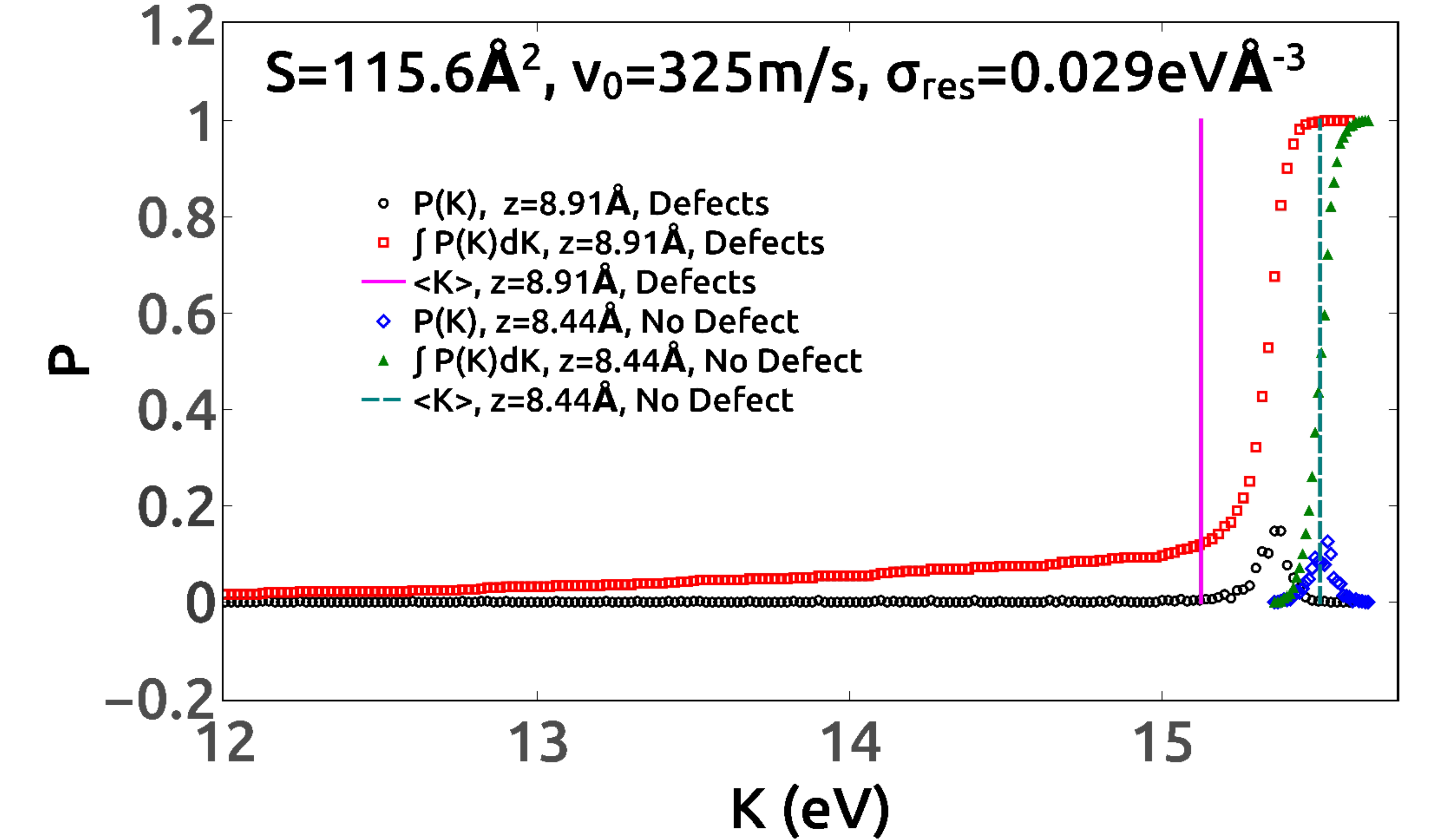}
\end{center}
\caption{
Kinetic energy distribution $P(K)$ shown for comparable $z$ positions during the retraction phase both with and without MgO-vacancy defects. Values of $P(K)$ shown for $z=8.91\AA$ with MgO-vacancy defects (open black circles) and $z=8.44\AA$ without defects (open blue diamonds). Running integrals $\int P(K) dK$  are shown for systems with MgO-vacancy defects (open red squares) and without defects (closed green triangles). Average values $\langle K \rangle$ are also shown for slabs with MgO-vacancy defects (solid magenta line) and without defects (dotted blue line).}
\label{figure13}
\end{figure}

Direct comparison of kinetic energy distributions for cases with and without MgO-vacancy defects, but otherwise identical DCS parameters, shows the connection between long-tails in the distribution, strong dissipation, and the presence of MgO-vacancy defects. In particular, Fig. \ref{figure13} shows a comparison between distributions. The $z$ coordinate for the two systems differs because of the previously noted difference in the starting separation due to the fact that the center-of-mass is somewhat displaced by removing MgO units from opposing surfaces. Therefore, the two $z$ coordinates used in Fig. \ref{figure13} actually correspond to the point where the restoring force has performed the same quantity of work on the system. Also notice that the $z$ coordinate represents slabs in the retraction phase which have moved far past their starting separation $z_{0}$. What is evident from the data in Fig. \ref{figure13} is that the ensemble with MgO-vacancy defects have a very long tail which includes $\sim 20\%$ of the ensemble within the tail, which extends to as much as $3$eV below the peak in $P(K)$. By contrast, the ensemble without MgO-vacancy defects exhibits a much more narrow distribution which, while not exactly a normal distribution, does have median and mean values which are reasonably close. The values $\langle K \rangle$ are also shown for the two ensembles. These demonstrate that the presence of MgO-vacancy defects is correlated with significantly stronger dissipation.

\begin{figure}
\begin{center}
\includegraphics[scale=0.25]{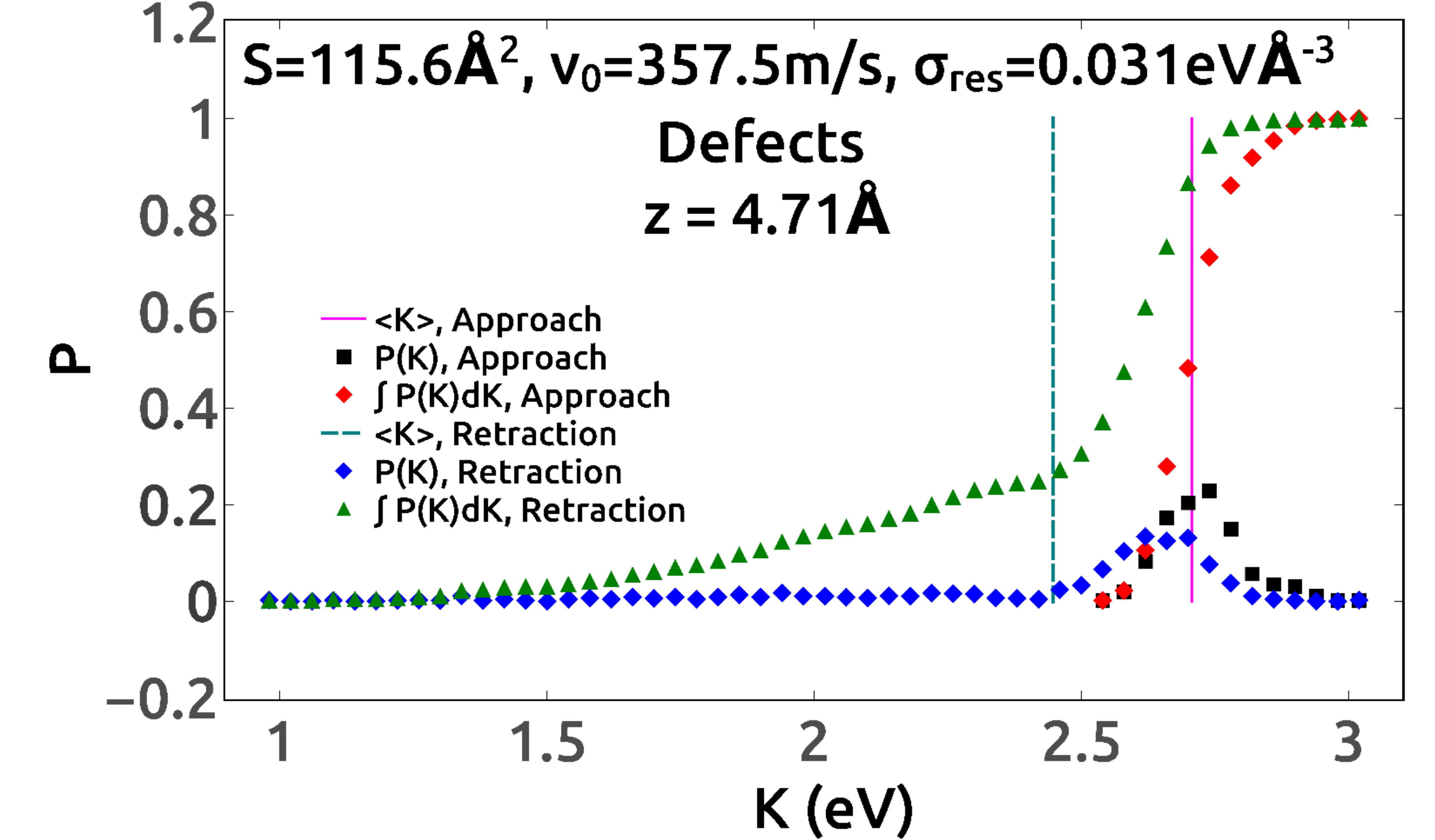}
\end{center}
\caption{ Kinetic energy distribution $P(K)$ for an ensemble of simulations where sticking events occurred. Not shown are the ensemble members which did stick, for which the final translational kinetic energy is zero. Results for $P(K)$ are shown for the approach (open black circles) and retraction (open blue diamonds) phases are shown at $z=4.71\AA$. Running integrals are shown for approach (closed red diamonds) and retraction (closed green triangles). The long tail in the distribution appears during retraction and is connected to strong ion rearrangement and even mass transfer between the slabs. Average values $\langle K \rangle$ are shown for approach (solid magenta line) and retraction (dashed green line).
}
\label{figure14}
\end{figure}

It is also possible to choose DCS parameters that result in dramatically larger dissipation and even the potential for sticking events, where the restoring stress $\sigma_{res}$ is insufficient to pull apart the interface. In particular an ensemble of N=500 simulations was computed including MgO-vacancy defects with $S=115.6\AA^{2}$, $v_{0}=357.5$ms$^{-1}$, and $\sigma_{res} = 0.031$eV$\AA^{-3}$. With these DCS parameters,  a total number $N_{stick}=45$ members of the ensemble resulted in sticking between the two slabs, for a probability $P_{stick}=0.088$. For the remaining calculations which do not yield sticking events, the probability distribution $P(K)$ is shown in Fig. \ref{figure14} at $z=4.71 \AA$. The tail in the distribution at lower kinetic energy values is substantially more pronounced than other calculations. Finally, because of the presence of sticking events in the ensemble, the JE was not applied to compute $\Delta A(z)$ and $W_{diss}(z)$. This is because the JE involves the calculation of free-energy differences between two equilibrium states, but  simulations with sticking events indicate the presence of two distinct final states (full adhesion and retracted surfaces) within the ensemble. Application of the JE to this situation is not clear.

In two of the simulation cases, both with MgO-vacancy defects, mass transfer between the slabs was found to occur in some members of the ensemble. Specifically, for small systems with $S=115.6\AA^{2}$, and DCS parameters $\sigma_{res}=0.029$eV$\AA^{3}$ and $v_{0}=325$ms$^{-1}$, there were $10$ ensemble members out of $N=500$ in which ions were swapped between the slabs. In this case, only Mg and O ions were involved, including transfer of a single Mg ion in 3 instances, and a single O ion in 7 instances. Therefore, the slabs acquire a small charge which changes the range and nature of the interactions. For DCS parameters $v_{0}=357.5$ms$^{-1}$, and $\sigma_{res} = 0.031$eV$\AA^{-3}$, which as we previously noted also resulted in some ensemble members sticking, a total of $38$ ensemble members exhibited ion swaps between the two slabs. Some of these involve single Mg or single O swaps, but a few included multiple ions and even some instances where $Si$ ions were transferred. Specifically, there were $18$ instances of a single O transfer, $13$ instances of a single Mg transfer, with each case leaving a small charge on the slabs. There were more extensive transfers which lead to charged slabs, including 1 instance where two O ions were transferred, 1 instance of a single SiO transfer, and 1 instance of a single SiO$_3$ transfer. Each of these events leave the slabs with an excess charge of $\pm 1.2 |e|$. Charge-neutral transfers also include 2 instances of transfer of a single MgO, and 2 instances of transfer of a single SiO$_2$. 

There is a question about what effect might be observed if the slabs were initially started with $z>6\AA$. While interactions are weak beyond this distance, the pair potential has a cutoff of $12\AA$ and therefore some interactions are present. It is first important to remember that the slabs initially are uncharged and possess no net dipole moment, even when defects were added. Consequently, very long-range electrostatic effects are only important during the retraction phase, and apparently only when defects are present in the initial structure. In principle, the JE determines free-energy differences, and this should not depend on the starting separation $z$. However, it is possible that defective structures in particular could be initialized at a larger separation to see if some additional dissipation is obtained. It is reasonable to expect that, while some dissipation should be obtained at larger initial $z$, that the calculations presented here likely have determine most of the dissipation.

To summarize the effects of MgO vacancies, at high enough defect concentrations the interaction between the slabs can result in disordering, mass transfer, and in some instances sticking. By contrast, slabs with lower MgO-vacancy density or with no defects do not exhibit mass transfer or sticking for the same DCS parameters. The displacement of ions during approach and especially retraction phases tends to lead to long-tail distributions in the kinetic energy, and hence strong dissipation. It appears that the relatively weaker binding of ions in the presence of vacancies allows ions to be more readily displaced from their initial equilibrium positions. In the final section of the article, we will discuss some of these observations in relation to stick-slip friction mechanisms, specifically where ions at an interface can become stuck in metastable states until the applied stress is sufficient to activate them to a new local minimum. 


\section{Discussion and Conclusions}

In this paper we have shown how ensemble calculations using the DCS approach along with the JE can reveal both the reversible surface free energy curve $\Delta A(z)/S$ and dissipative properties. In contrast to previous applications of the JE, the approach described here enables calculation of the interfacial dissipation $\gamma_{diss}(z)$ at all points along the trajectory. This suggests an approach which can link discrete atomic-scale events to dissipation. In the simulations reported here, the approach reveals several surprising and interesting aspects of dissipation during a DCS simulation.

\begin{figure}
\begin{center}
\includegraphics[scale=0.20]{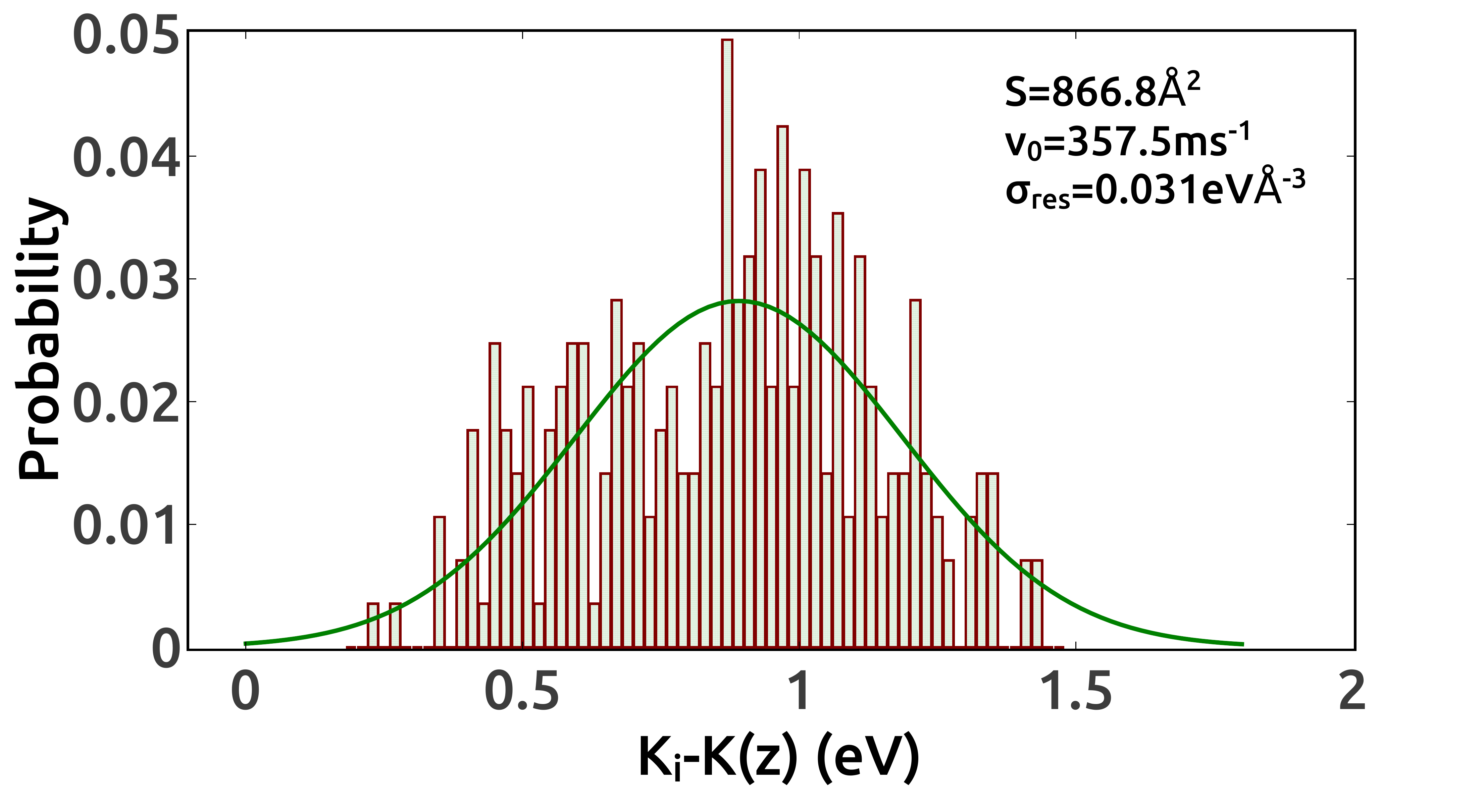}
\end{center}
\caption{Probability distribution for $K_{i}-K(z)$ at $z=z_{0}$ during the retraction phase. This is a typical distribution with multiple peaks consistent with far-from-equilibrium behavior. The DCS parameters are shown in the figure, and the ensemble consisted of surfaces without MgO-vacancy defects. The curve represents the best Gaussian fit to the data.
}
\label{figure15}
\end{figure}

First, it is apparent that for all of the results reported here, the conditions can best be described as consistent with a far-from-equilibrium regime. Specifically, kinetic-energy distributions are never found to be described by a Gaussian, and in cases of very strong dissipation exhibit long-tailed distributions. In Fig. \ref{figure15} the probability distribution for $K_{i}-K(z)$ at $z=z_{0}$ during retraction is shown for an ensemble without MgO-vacancy defects. This distribution is clearly not Gaussian, but is instead characterized by multiple peak values. The distribution is representative of the general behavior observed. Namely, while lacking the long-tailed distribution seen when MgO vacancies are present, the distribution nevertheless appears consistent with far-from-equilibrium behavior. 

In the case of surfaces with MgO-vacancy defects, dramatically different behavior was observed. Namely, ions at defective surfaces can be more easily dislodged during close approach of two flat surfaces. When this occurs, mass transfer is possible, with an additional disordering of the surfaces. These cases are closely associated with even stronger dissipation and long tails at very low kinetic energy during the rebound phase. Specifically, while average dissipation tends to be in many cases below $1\%$ of the incident kinetic energy, slabs with defects exhibit dissipation up to $\sim 3\%$. Moreover, approach distances are not less than $\sim 3 \AA$, indicating much stronger dissipation is possible. However, even for closest approach distances slightly greater than $\sim 3\AA$, many members of the ensemble exhibit dissipation of in the range $20-30\%$ of their incident kinetic energy.  Closer approach distances are shown to result in very strong dissipation and even sticking between the surfaces, where the restoring force is unable to separate the slabs. 

\begin{figure}
\begin{center}
\includegraphics[scale=0.20]{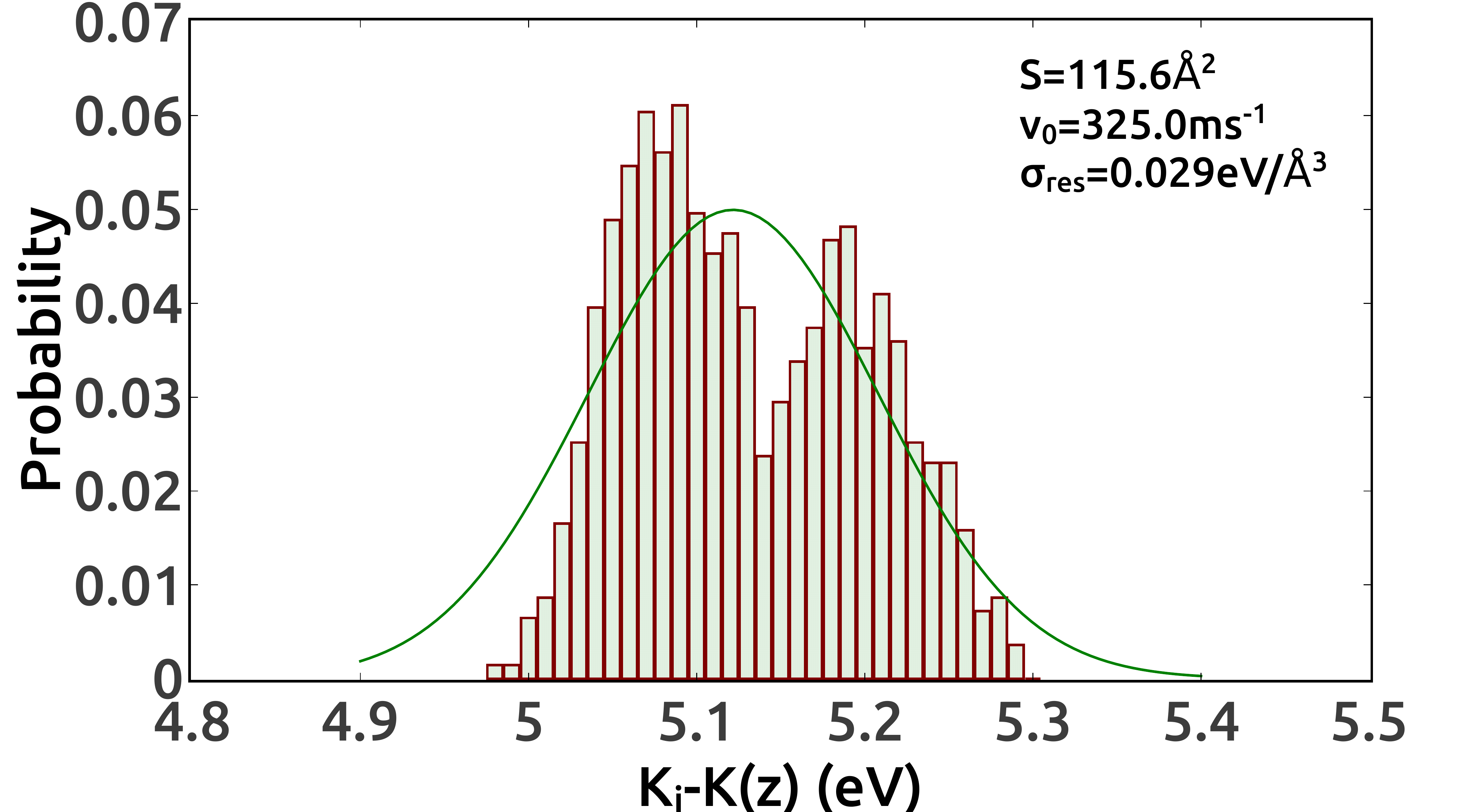}
\end{center}
\caption{Probability distribution for $K_{i}-K(z)$ at $z=4.45\AA$ during the retraction phase. This is a typical distribution with multiple peaks consistent with far-from-equilibrium behavior. The DCS parameters are shown in the figure, and the ensemble consisted of surfaces without MgO-vacancy defects. The curve represents the best Gaussian fit to the data.
}
\label{figure16}
\end{figure}

Strong dissipation appears to be closely linked to atoms at the very outermost layer which can potentially be displaced or even entirely removed from their local harmonic well by the approach of the opposing surface. This is similar to the ``stick-slip'' mechanism which has been used to describe nanoscale sliding friction experiments\cite{Socoliuc:2004wx,KRIM2002741}. Specifically, stress can build during the motion of two opposing surfaces until interfacial bonds become activated. When bonds break, the slipping can occur, which is characterized by motion for some time with very low dissipation\cite{Socoliuc:2004wx}. We propose that a very similar process occurs in the DCS calculations reported here. Namely, on approach, bonds become stretched due to interactions between the two surfaces, and new local minima will develop. During retraction, stress builds until ions are activated and bonds across the interface abruptly break. This is consistent with the JKR mechanism. Because the local events relate to individual bond-breaking, this likely leads to far-from equilibrium behavior and the non-Gaussian distributions like Fig. \ref{figure15} and Fig \ref{figure16}. Figure \ref{figure16} appears to show a distinctly bimodal distribution. While Fig. \ref{figure15} is more complex since it shows a much broader range for $K_{i}-K(z)$ while also corresponding to a smaller ensemble $N=200$, simple statistical analysis confirms the deviations from the Gaussian fits are not simply insufficient sampling. Specifically, each bin should be governed by a binomial distribution, with probability $p$ a given ensemble member falls in any given bin, and probability $1-p$ it does not. Applying the central-limit theorem to each bin, the standard deviation is $\sigma \approx \sqrt{p(1-p) \over N}$, where $N$ is the total number of members of the ensemble. For Fig. \ref{figure15}, $N=200$, and then with $p=0.03$ near the center of the distribution, $\sigma \approx 0.012$.  The data in Fig. \ref{figure15} shows many points with deviations from the Gaussian fit that are substantially larger than this estimate. Similarly for Fig. \ref{figure16}, $N=500$, and $p=0.05$ near the center of the distribution. The same estimate yields $\sigma \approx 0.0097$, which is smaller than deviations from the Gaussian fit, and smaller than the apparent systematic deviations that appear to be a bimodal distribution. Based on this analysis, the distributions clearly deviate strongly from a single normal distribution.

The substantial deviation for normal distributions in Fig. \ref{figure15} and Fig. \ref{figure16} appear consistent with the idea that the physics is governed by discrete atomic-scale events associated with activated processes. When MgO-vacancy defects are present, ions are more easily displaced by the approaching surfaces, leading to even larger dissipation and even mass transfer between the slabs. 

When substantial disruptions occur, which is generally observed for separations below $\sim 3\AA$ in the case of defective forsterite surfaces, the process can lead to total dissipation and sticking. However, even small numbers of these disruptive events leads to large tails in the kinetic energy distribution, even when there are often a large number of ensemble members with low dissipation or even negative dissipation, and hence apparently good averaging of the JE.

The finding that $\gamma_{diss}(z)$ decreases as the slab separation $z$ increases is a very surprising finding. The mechanism for this behavior is not entirely certain, but it is possible to speculate that it also arises due to stick-slip type behavior. Specifically, in typical stick-slip motion related to sliding friction, stress builds until a discrete event occurs, at which point interfaces can briefly slide with very low dissipation\cite{Socoliuc:2004wx}. In the simulations reported here, it is certain that these abrupt bond-breaking events occur at different times for each member of the ensemble. Perhaps this leads to a kinetic energy distribution that initially widens as some ensemble members ``slip'' before others. However,  other members will simply enter the low-dissipation ``slip'' phase at a later time. This could potentially narrow the kinetic energy distribution and hence correspond to decreasing $\gamma_{diss}(z)$.

Another possible explanation for decreasing $\gamma_{diss}(z)$ with increasing $z$ during retraction is that it is simply due to errors in evaluating the Jarzynski estimator. We have only attempted to evaluate the error at $z=z_{0}$ after retraction. Certainly at $z=z_{0}$, the entirety of the difference between $\gamma_{diss}(z_{0})$ and $-\gamma_{int}$ is due to statistical bias in $\Delta A$ reflected in Table \ref{table2} and Table \ref{table3}.  However, the error is significantly smaller than the magnitude of the decrease in $\gamma_{diss}(z)$ during retraction. Yet we expect that the statistical bias varies along the trajectory and hence  the error in computing $\gamma_{diss}(z)$ could depend on $z$.

While some of the above discussion is somewhat speculative, it is nonetheless clear that dissipation is associated with discrete atomic-scale events where bonds are formed and later broken under stress. While the JE is still useful in this regime, it is challenging to obtain results that are not subject to biasing and numerical issues encountered even in near-equilibrium calculations. Specifically, the tendency of $\Delta A(z)$ to be biased to more positive values corresponds to an underestimation of $\gamma_{diss}(z)$. Previous work indicates that convergence and elimination of bias might only occur gradually with increasing the number of ensemble members\cite{Gore:2003vb}. However, previous studies have focussed on near-equilibrium cases which does not correspond to the regime simulated in this paper. Moreover, certain fundamental questions exist with respect to free energy differences between quasi-equilibrium systems\cite{Williams_2008}. The potential for quasi-equilibrium ensembles in the results here are indicated by the presence of local energy minima which are especially evident when MgO-vacancy defects are present. In systems which can readily explore neighboring free-energy minima, the potential for strong dissipation is evident.

Finally, the use of the JE as a way to quantify dissipation might also enable more accurate computational methodology. Specifically, because dissipation can be elucidated without requiring a complete approach and retraction phase (i.e. a closed cycle), it might be possible to examine dissipation for small surfaces modeled using accurate density-functional theory (DFT) methodology. Moreover, the present work suggests that exploration of activated behavior at interfaces might be a useful direction to explore in the context of dissipation. This could involve examination of interfacial reaction mechanisms perhaps using standard approaches to estimate reaction barriers. Because the JE allows both $\Delta A(z)$ and dissipative work to be computed while atomic rearrangements take place, it might be used to specifically identify the point along a trajectory and the detailed mechanism associated with the largest dissipative effect.

\section{Acknowledgments}

This research was made possible by support from the National Science Foundation, Division of Astronomical Sciences, USA, under grant number 1616511. We also acknowledge support from the Advanced Research Computing Center at the University of Central Florida, USA for use of the STOKES computing cluster.

\section{Data availability}
The data that support the findings of this study are available from the corresponding author upon reasonable request.
\newpage

\bibliographystyle{unsrt}

\end{document}